\RequirePackage[l2tabu,orthodox]{nag}

\newif{\ifiopformat}
\iopformatfalse

\newif{\ifhavepgf}
\havepgffalse

\ifiopformat
\documentclass[12pt,a4paper]{iopart}
\else
\documentclass[aps,prb,reprint,showpacs,showkeys,floatfix,a4paper]{revtex4-1}
\fi

\usepackage[utf8]{inputenc}
\usepackage[english]{babel}
\usepackage{csquotes}

\ifiopformat
\usepackage{iopams}
\expandafter\let\csname equation*\endcsname\relax
\expandafter\let\csname endequation*\endcsname\relax
\usepackage[numbers,square,sort&compress]{natbib}
\newcommand{\refcite}[1]{\cite{#1}}
\newcommand{\refscite}[1]{\cite{#1}}
\else
\providecommand{\submitto}{}
\providecommand{\SUST}{}
\usepackage{natbib}
\newcommand{\refcite}[1]{Ref. \onlinecite{#1}}
\newcommand{\refscite}[1]{Refs. \onlinecite{#1}}

\fi
\usepackage{natmove} %

\usepackage{amssymb}
\usepackage{mathtools}
\usepackage{commath}
\usepackage{physics}
\usepackage{bm}
\usepackage[mode=math,range-units=single]{siunitx}
\newcommand{\mat}[1]{\ensuremath{\mathsf{#1}}}
\newcommand{\vc}[1]{\ensuremath{\vb{#1}}}
\renewcommand{\Re}[1]{\ensuremath{\real\Bqty{#1}}}
\renewcommand{\Im}[1]{\ensuremath{\imaginary\Bqty{#1}}}

\usepackage{xcolor}
\usepackage[caption=false]{subfig}

\ifhavepgf
\usepackage{pgfplots}
\usepackage{tikz}
\usetikzlibrary{plotmarks}
\usetikzlibrary{matrix}
\usetikzlibrary{calc}
\usetikzlibrary{external}
\pgfplotsset{compat=1.11}
\pgfplotsset{plot coordinates/math parser=false}
\tikzsetexternalprefix{tikz/}
\usetikzlibrary{shapes}
\usetikzlibrary{decorations.markings}
\usetikzlibrary{positioning}
\usetikzlibrary{backgrounds}
\usetikzlibrary{spy}
\usepackage[european,cuteinductor,siunitx,smartlabels]{circuitikz}

\tikzset{%
  /tikz/external/only named = true,
  /tikz/external/optimize = false, %
}

\tikzset{%
  every pin edge/.append style = {<-, semithick, black},
  every pin/.append style = {pin distance=3mm, font=\scriptsize, black},
  subplotleft/.style = {baseline,
    trim right={($(current axis.south east) + (0.75em,0)$)},
    trim left={($(current axis.south west) - (3em,0)$)},
    every node/.prefix style = {inner sep=0.3333em}},
  subplotright/.style = {baseline,
    trim right={($(current axis.south east) + (0.75em,0)$)},
    trim left={($(current axis.south west) - (2.5em,0)$)},
    every node/.prefix style = {inner sep=0.3333em}},
  plotcontainer/.style = {inner sep=0, anchor=base east},
}

\pgfkeys{%
  /pgf/number format/set thousands separator={\,},
}
\pgfplotsset{%
  xlabel near ticks,
  ylabel near ticks,
  every axis plot post/.append style = semithick,
  axis line style = thin,
  tick style = thin,
  subplotaxis/.style = {
    tick label style = {font=\footnotesize},
    label style = {font=\small}, %
    tick label style = {inner xsep=0.125em},
    label style = {inner xsep=0.125em},
  },
}

\else

\usepackage{graphicx}
\usepackage{tikzexternal}
\tikzsetexternalprefix{tikz/}

\fi

\newlength{\figureheight}
\newlength{\figurewidth}
\ifiopformat
\else
\fi
\usepackage{ifpdf}
\ifpdf
\usepackage[pdftex]{hyperref} %
\fi
\usepackage[noabbrev]{cleveref} %
\crefname{equation}{\unskip}{\unskip}
\Crefname{equation}{Equation}{Equations}
\usepackage{microtype} %

\usepackage{array}
\usepackage{booktabs}

\ifhavepgf
\usepackage{pgfplotstable}
\pgfplotstableset{%
  include outfiles,
  every table/.append style={outfile={tikz/#1.tex}}, %
  every head row/.style = {before row=\toprule, after row=\midrule},
  every last row/.style = {after row=\bottomrule},
  row style/.style 2 args={
      every row #1 column 0/.style={#2},
      every row #1 column 1/.style={#2},
      every row #1 column 2/.style={#2},
      every row #1 column 3/.style={#2},
      every row #1 column 4/.style={#2},
      every row #1 column 5/.style={#2},
      every row #1 column 6/.style={#2},
      every row #1 column 7/.style={#2},
      every row #1 column 8/.style={#2},
      every row #1 column 9/.style={#2},
      every row #1 column 10/.style={#2},
  }
}
\else

\fi

\newcommand{\mytitle}{Electrothermal Feedback in Kinetic Inductance Detectors}
\newcommand{\tu}[1]{\ensuremath{_{\mathrm{#1}}}}
\newcommand{\ts}[1]{\ensuremath{^{\mathrm{#1}}}}

\newcommand{\SiN}{\ensuremath{\text{Si}_{x}\text{N}_{y}}}
\newcommand{\imi}{\ensuremath{\mathrm{i}}}

\ifpdf
\hypersetup{%
  pdfinfo={%
    Title={\mytitle},
    Author={T Guruswamy, C N Thomas, S Withington and D J Goldie},
  }
}
\fi

\begin{document}
\title[]{\mytitle}
\ifiopformat
\author{T Guruswamy, C N Thomas, S Withington and D J Goldie}
\ead{tg307@mrao.cam.ac.uk}
\address{Quantum Sensors Group, Cavendish Laboratory, University of Cambridge,
J J Thomson Avenue, Cambridge, CB3~0HE, UK}
\else
\author{T Guruswamy}
\email{tg307@mrao.cam.ac.uk}
\author{C N Thomas}
\author{S Withington}
\author{D J Goldie}
\affiliation{Quantum Sensors Group, Cavendish Laboratory, University of Cambridge,
J J Thomson Avenue, Cambridge, CB3~0HE, UK}
\fi
\date{\today}

\begin{abstract}
In Kinetic Inductance Detectors (KIDs) and other similar applications of superconducting microresonators, both the large and small-signal behaviour of the device may be affected by electrothermal feedback. Microwave power applied to read out the device is absorbed by and heats the superconductor quasiparticles, changing the superconductor conductivity and hence the readout power absorbed in a positive or negative feedback loop. In this work, we explore numerically the implications of an extensible theoretical model of a generic superconducting microresonator device for a typical KID, incorporating recent work on the power flow between superconductor quasiparticles and phonons. This model calculates the large-signal (changes in operating point) and small-signal behaviour of a device, allowing us to determine the effect of electrothermal feedback on device responsivity and noise characteristics under various operating conditions. We also investigate how thermally isolating the device 
from the bath, for example by designing the device on a membrane only connected to the bulk substrate by thin legs, affects device performance. We find that at a typical device operating point, 
positive electrothermal feedback reduces the effective thermal conductance from the superconductor quasiparticles to the bath, and so increases responsivity to signal (pair-breaking) power, increases noise from temperature fluctuations, and decreases the Noise Equivalent Power (NEP). Similarly, increasing the thermal isolation of the device while keeping the quasiparticle temperature constant decreases the NEP, but also decreases the device response bandwidth.
\end{abstract}
\submitto{\SUST}
\maketitle

\section{Introduction}
Kinetic Inductance Detectors (KIDs) are ultrasensitive photon detectors based on superconducting microresonators~\cite{Day2003}. Their low noise and ease of multiplexing leads to a wide variety of applications, particularly in astrophysics~\cite{Vardulakis2008} and particle physics~\cite{Cruciani2016}. The principles of operation are very similar to those used in superconducting qubits~\cite{DiCarlo2010,Hofheinz2008,Schoelkopf2008} %
and superconducting quantum interference device (SQUID) multiplexers~\cite{Irwin2004}.
During the normal operation of a KID, absorbed signal power breaks Cooper pairs in the superconductor, changing the quasiparticle effective temperature. This causes a change in the superconductor impedance and so a shift in the amplitude and phase of a microwave readout tone transmitted through the device.

Most existing models of KIDs~\cite{Zmuidzinas2012} do not include a detailed analysis of how power absorbed by the superconductor from the readout tone affects the quasiparticle temperature, or the rate at which the quasiparticles cool via the superconductor phonons. Recent work~\cite{DeVisser2010,Thompson2013} has shown that both in theory and experiment, quasiparticle heating due to readout power is significant under typical device operating conditions, and leads to nonlinear device behaviour.
Furthermore, detailed calculations of quasiparticle-photon and quasiparticle-phonon interactions in superconductors~\cite{Goldie2013,Guruswamy2015b} have shown the quasiparticle-phonon power flow in a superconductor has a significantly different functional form to the normal state electron-phonon power flow expressions typically used, and also changes with readout power, frequency, and bandwidth~\cite{Guruswamy2016}. Incorporating these findings in a higher level device model, \textcite{Thomas2014} described a extensible framework which combines quasiparticle heating with the electrical behaviour of the resonator and allows calculation of experimentally relevant measurements.
In this work, we explore numerically the implications of the theoretical model for a typical device configuration: an Al thin film resonator with resonant frequency \SI{5}{GHz} at a bath temperature of $0.1\,T_c = \SI{0.118}{K}$, where $T_c$ is the superconducting critical temperature of Al.

In particular, as the model of \refcite{Thomas2014} and the implementation described here include both readout power heating and superconductor quasiparticle-phonon cooling, we can calculate the effect of \emph{electrothermal feedback} on device performance. Here we define electrothermal feedback as the phenomenon where readout power heats the superconductor quasiparticles of a superconducting microresonator, changing the superconductor impedance, and hence the readout power absorbed. In principle, electrothermal feedback could affect both the magnitude and bandwidth of the device response to signal power, and noise, and could be positive or negative depending on the details of the readout frequency, power, and resonator characteristics.
In this work we quantify the effect of electrothermal feedback on device responsivity and noise for a typical device configuration as in \textcite{DeVisser2010}%

We also investigate the effect on device performance of intentionally varying the thermal isolation of the device from the bath, which may be expected to change the device dynamic range, responsivity, noise characteristics, or even to control electrothermal feedback and onset of nonlinearities due to readout power heating. In practice, the thermal isolation could be controlled by depositing the superconducting resonator on a membrane only connected to the bulk substrate by thin legs.
This principle is already standard for other types of photon detectors such as Transition Edge Sensors~\cite{Goldie2011,Osman2014}, and is used in recent KID designs~\cite{Quaranta2013,Timofeev2014,Lindeman2014b}. As the context of this work is using superconducting resonators as detectors, we investigate the effects of this thermal isolation on responsivity to signal power and noise.

The model represents the device with an equivalent series electrical circuit and a thermal model of several connected heat capacities with a number of external power sources, dependent both on the temperatures of each element, and a set of externally controlled parameters. This framework allows the steady-state operating point (set of temperatures and readout output) and the small-signal behaviour about that operating point to be determined. This small-signal analysis allows us to define and calculate the responsivity to signal (pair-breaking photons) power, required for operating these devices as detectors. We also follow \refcite{Thomas2014} in introducing noise sources based on temperature fluctuations, and calculate the resulting Noise Equivalent Power (NEP) as device operating point, electrothermal feedback, and thermal isolation are varied.
Other models investigating electrothermal feedback based on bolometer theory exist~\cite{Lindeman2013,Lindeman2014}, but we consider this work more general as we are able to simultaneously consider the large and small-signal behaviour of the device, and include more complex thermal configurations beyond a single thermal link between the superconductor and bath. We are therefore able to consider bolometric operating modes, a subject of ongoing work~\cite{Timofeev2014,Cardani2015}.
In \cref{sec:methods} we describe the model in further detail and its numerical implementation for a typical KID design. \Cref{sec:results} discusses the dependence of the experimentally relevant measurement outputs on key parameters, in particular focussing on the key questions of how electrothermal feedback and varying the thermal isolation of the device affect device responsivity and noise characteristics.

\section{Methods} \label{sec:methods}
\subsection{Electrical circuit} \label{sec:electrical}
We represent the electrical behaviour of the device using a series equivalent circuit as illustrated in \textcite[figure 3]{Thomas2014}.
A superconducting film (resistance $R$ and inductance $L$) is coupled by a capacitance $C$ to a driving generator and readout load of impedance $Z_0$.

One of the goals of this work is to be as independent of device geometry as possible. We therefore parametrise the circuit in terms of a resonance frequency $\nu_0 = 1/2\pi\sqrt{LC}$, internal quality factor $Q_I = 2\pi\nu_0 L/R$, and coupling quality factor $Q_C = 4\pi\nu_0 L/Z_0$.
Instead of choosing a device geometry and calculating $R$, $L$, and $C$, we specify the generator and load impedance $Z_0$, resonator frequency $\nu_0$ and quality factors $Q_I$ and $Q_C$ at resonance and when the superconductor quasiparticle temperature $T_{qp}$ is equal to the bath temperature $T_b$. This uniquely specifies $R(T_b)$, $L(T_b)$, and $C$.

In this electrical equivalent circuit, the resistance $R$ is attributed entirely to quasiparticle losses within the superconductor, calculated from the real part of the surface impedance $Z_s$.
However the inductance $L$ is the sum of the magnetic inductance $L\tu{geo}$ due to the geometry of the film, as well as that due to the imaginary part of the superconductor surface impedance, $L\tu{s}$. As the quasiparticle temperature changes from $T_b$ to $T_{qp}$, we modify the values of $R$ and $L$ as functions of readout frequency $\omega = 2\pi\nu_r$ as
\begin{align}
  R(\omega, T_{qp}) &= R(\omega, T_b) \frac{\Re{Z_s(\omega, T_{qp})}}{\Re{Z_s(\omega, T_b)}} \;, \\
  L(\omega, T_{qp}) &= L(\omega, T_b) + L_{s}(\omega, T_b)\left( \frac{\Im{ Z_{s}(\omega, T_{qp})}}{ \Im{ Z_{s}(\omega, T_b)} } - 1 \right) \; .
\end{align}
For the purposes of this paper, both $R$ and $L_s$ are related to the surface impedance by the same geometric factor~\cite{Gao2008a}, and so for consistency we require
\begin{equation}
  L_s(\omega, T_b) = \frac{1}{\omega} R(\omega, T_b) \frac{\Im{ Z_s(\omega, T_b) }}{\Re{ Z_s(\omega, T_b) }} \label{eq:kinetic_inductance_fraction} \; .
\end{equation}
Defining the kinetic inductance fraction $\alpha = L_s(\omega, T_{qp})/L(\omega, T_{qp})$, \cref{eq:kinetic_inductance_fraction} is equivalent to the known expression for internal quality factor~\cite{Zmuidzinas2012}
\begin{equation}
  \frac{1}{\alpha}\frac{\Im{ Z_s(\omega, T_{qp}) }}{\Re{ Z_s(\omega, T_{qp}) }} = Q_I \; . \label{eq:qi_condition}
\end{equation}
Considered as an inequality ($\alpha \le 1$), \cref{eq:qi_condition} gives the lower limit of $Q_I$ possible in a model with only quasiparticle losses.

The surface impedance $Z_s$ of a superconducting film is calculated in the local, thick film limit as
\begin{equation}
  Z_s(\omega, T_{qp}) = \sqrt{\frac{\imi\omega\mu_0}{\sigma(\omega, T_{qp})}}
\end{equation}
where $\sigma(\omega, T_{qp}) = \sigma_1(\omega, T_{qp}) - \imi\sigma_2(\omega, T_{qp})$ is the complex conductivity calculated using the Mattis-Bardeen equations~\cite{Mattis1958}. The specific form of the surface impedance expression is not crucial to the model; the thin film regime ($Z_s \propto \sigma^{-1}$) could equally be considered~\cite{Zmuidzinas2012}.

For readout frequencies $\nu_r$ close to the resonance frequency $\nu_0$, the de-embedded values of the transmission ($S_{21}$) and reflection ($S_{11}$) $S$-parameters, as measured at the device plane, are~\cite{Thomas2014}
\begin{align}
  S_{11}(\nu_r, T_{qp}) &= -\frac{Q_T}{Q_C}\left( 1 + 2 \imi Q_T \frac{\nu_r - \nu_0}{\nu_0} \right)^{-1} \;, \\
  S_{21}(\nu_r, T_{qp}) &= 1 + S_{11}(\nu_r, T_{qp}) \;,
\end{align}
where $\nu_r$ is the readout frequency, total quality factor $Q_T = Q_I Q_C/(Q_I + Q_C)$, and we emphasise that all of $\nu_0$, $Q_I$, and $Q_C$ are dependent on quasiparticle temperature $T_{qp}$.
We are able to calculate the scattering parameters as functions of detuning of the readout frequency $\Delta\nu = \nu_r - \nu_0$ and quasiparticle temperature $T_{qp}$, initialising the calculations with only the device parameters $Z_0$, $\nu_0$, and quality factors $Q_I$ and $Q_C$ at $T_b$, as well as relevant material properties, but with no dependence on a specific device geometry.
In test calculations of $S_{21}$ as a function of readout frequency, as the quasiparticle temperature is increased, the resonance peak shifts to a lower frequency and becomes shallower and broader in agreement with detailed microstrip-based models.
As the readout power absorbed by the resonator $q_R$ is
\begin{equation}
  q_R = P_R\left(1 - \abs{S_{21}}^2 - \abs{S_{11}}^2\right) \; \,
\end{equation}
for an incident readout power $P_R$, we are now able to proceed to a full electrothermal model.

\subsection{Thermal configuration} \label{sec:thermal}

We model the thermal behaviour of a generic KID with three heat capacities: the superconductor quasiparticles, the superconductor phonons, and the substrate phonons, as shown in the block diagram inset of \cref{fig:1}. Each has its own effective temperature: $T_{qp}$, $T_{ph}$, and $T_{su}$ respectively.
The heat capacity of the superconductor quasiparticles is given by BCS theory~\cite{Tinkham2004}; while the heat capacity of the superconductor phonons are assumed to follow a Debye model. The substrate material of \SiN{}, an amorphous dielectric, is known to be dominated by two-level systems and has a heat capacity linear in temperature~\cite{Phillips1987} such that $C_{su} = c\,V_{su}\,T_{su}$, where $V_{su}$ is the total volume of the substrate, and $c = \SI{26.44}{J.m^{-3}.K^{-2}}$ is an empirically derived constant for \SiN{}~\cite{Rostem2008a}.
Three possible external sources of power are included. The device readout, at frequencies below the superconducting gap frequency, heating the quasiparticles ($q_R$); a signal, of photons at frequencies above the gap frequency, also heating the quasiparticles ($q_S$); and any power heating the substrate upon which the device is patterned ($q_H$).

The readout power absorbed $q_R$ has been described in \cref{sec:electrical}.
Power flow between the superconductor quasiparticles and phonons $q_I$ is described by a two-temperature exponential approximation to the total power flow~\cite{Goldie2013,Guruswamy2015b}. In general, superconductor quasiparticles and phonons interact both by inelastic scattering and by quasiparticle recombination into Cooper pairs. By appropriately integrating the Chang \& Scalapino kinetic equations~\cite{Chang1977,Chang1978} describing the rate of change of the quasiparticle energy distribution due to these interactions, we derived the net power flow due to recombination $q_{rec}$, and the total power flow $q_I$, when $T_{qp} \ll T_c$ and $T_{ph} < T_{qp}$ in clean superconductors as
\begin{align}
  \begin{split}
    q_{rec} &= V \Sigma_{s} \left( T_{qp} \exp(-2\Delta(T_{qp})/k_B T_{qp}) \right. \\
    &\qquad \left. - T_{ph} \exp(-2\Delta(T_{ph})/ k_B T_{ph}) \right)
  \end{split} \label{eq:recombination} \\
  q_{I} &= q_{rec} / \eta_r  \; , \label{eq:super_q}
\end{align}
where $V$ is the volume of the superconductor, $\Sigma_s$ is a material-dependent constant, in this case $\SI{3.23E10}{W.m^{-3}.K^{-1}}$ for Al~\cite{Guruswamy2015b}, and $\Delta(T)$ is the superconducting gap energy at temperature $T$. This expression for $q_{rec}$ has been explored both in understanding nonequilibrium behaviour in resonators~\cite{Goldie2013,Guruswamy2015b} and in superconducting tunnel junctions~\cite{Maisi2013}. The scattering portion of the power flow could also be described by an approximate expression for a quasiparticle distribution in quasiequilibrium~\cite{Maisi2013}, but we find after considering the nonequilibrium quasiparticle distributions resulting from heating by a single frequency microwave tone, it is better accounted for in this case by scaling $q_{rec}$ by a factor $1/\eta_r$, where $\eta_r$ is the power and microwave frequency dependent fraction of the total power flow carried by recombination, derivable from the full solutions of the nonequilibrium kinetic equations as outlined in \refscite{Goldie2013,Guruswamy2015b}. Here, readout frequency does not vary significantly, so only the power dependence of $\eta_r$ is retained. For Al and a \SI{5}{GHz} microwave readout, $\eta_r < 0.6$ and has a logarithmic dependence on power~\cite{Guruswamy2015b} above power densities of $\SI{2.83}{mW.m^{-3}}$.
The phonon trapping factor $\left(1+\tau_l/\tau_{pb}\right)$ introduced in \refcite{Goldie2013} is eliminated in favour of an explicit heat flow term between the superconductor phonons and substrate phonons.

\begin{figure}[htb]
  \centering
  \tikzsetnextfilename{el_figure_5_q_curves}
  \input{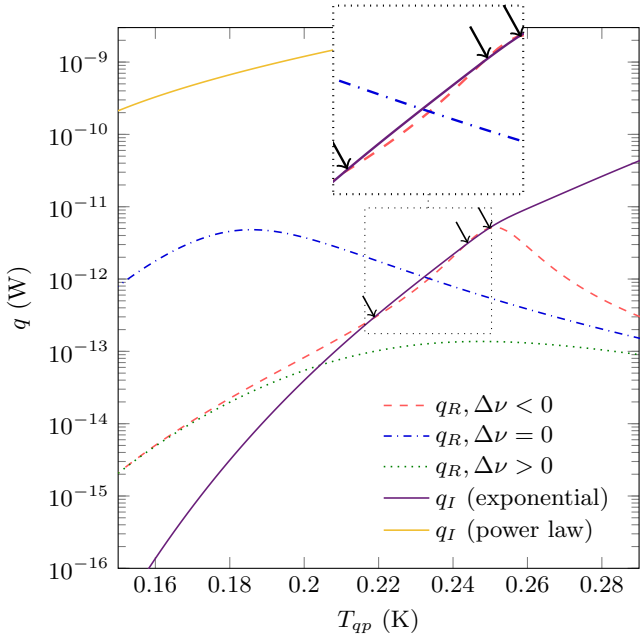}
  \caption{Example of readout power absorbed by superconductor quasiparticles, $q_R$, below resonant frequency ($\Delta\nu = \nu_r - \nu_0 < 0$), at resonant frequency ($\Delta\nu = 0$), and above resonant frequency ($\Delta\nu > 0$), all as a function of quasiparticle temperature $T_{qp}$. The power flow out of the superconductor quasiparticles and into the superconductor phonons, $q_I$ \cref{eq:super_q}, is included as a dashed line. For comparison the normal state power law expression \cref{eq:normal_q} is also included. Where $q_R$ and $q_I$ intersect, as indicated by the arrows, represents a possible steady-state operating point for the quasiparticles. Al resonator, with $P_R = \SI{20}{pW}$.}
  \label{fig:5}
\end{figure}

\begin{figure}[htb]
  \centering

  \ifhavepgf
  \tikzset{external/remake next}
  \tikzexternaldisable
  \newsavebox\mainplot
  \savebox\mainplot{
%
%
\begin{tikzpicture}

\begin{axis}[%
width=0.967\figurewidth,
height=\figureheight,
at={(0\figurewidth,0\figureheight)},
scale only axis,
xmode=log,
xmin=0.002,
xmax=6.32455532033676,
xlabel style={font=\color{white!15!black}},
xlabel={$g_{su}$},
ymode=log,
ymin=1e-10,
ymax=0.3,
ylabel style={font=\color{white!15!black}},
ylabel={$\tau$ (s)},
extra description/.code = {\node[coordinate,pin={above:{$\tau_1 \!\sim\! \dfrac{C_{ph}}{G_P}$}}] at (axis cs:2,1.533e-10) {}; \node[coordinate,pin=above:{$\tau_3 \!\sim\! C_{qp} / G_I$}] at (axis cs:0.01,0.0003652) {}; \node[coordinate,pin=above:{$\tau_2 \!\sim\! \Sigma C / G_L$}] at (axis cs:0.1697,0.003151) {};}extra description/.code = {  }extra description/.code = {  }
]
\addplot [color=white!35!red, forget plot]
  table[row sep=crcr]{%
0.002	0.275563200246786\\
0.00235753726958717	0.232673233812025\\
0.00277899098874627	0.196602705744252\\
0.00327578741390813	0.166226421466908\\
0.0038613954577665	0.140616415069177\\
0.00455169185214958	0.119004152053059\\
0.00536539159055945	0.100750850619674\\
0.00632455532033676	0.0853239485254926\\
0.00745518744062988	0.072278282010925\\
0.00878794112152158	0.0612409162913561\\
0.0103589493584624	0.0518988353927068\\
0.0122108045931707	0.0439888926387841\\
0.014393713460023	0.0372895629247119\\
0.0169668579648814	0.0316141427589866\\
0.02	0.0268051210932657\\
0.0235753726958717	0.0227295033792294\\
0.0277899098874628	0.0192749154952551\\
0.0327578741390813	0.0163463491014953\\
0.038613954577665	0.0138634367262072\\
0.0455169185214958	0.0117581658934371\\
0.0536539159055945	0.00997295845582796\\
0.0632455532033676	0.00845905443924934\\
0.0745518744062988	0.00717515050795893\\
0.0878794112152158	0.00608625172571216\\
0.103589493584624	0.00516270235242872\\
0.122108045931707	0.00437936714319599\\
0.14393713460023	0.00371493933324667\\
0.169668579648814	0.0031513553869284\\
0.2	0.00267329981040182\\
0.235753726958718	0.00226778600405669\\
0.277899098874627	0.0019238013610542\\
0.327578741390813	0.00163200668536885\\
0.38613954577665	0.00138448156388198\\
0.455169185214958	0.00117450864464556\\
0.536539159055945	0.000996390894712144\\
0.632455532033676	0.000845296895048305\\
0.745518744062988	0.000717130202020758\\
0.878794112152158	0.000608420212611525\\
1.03589493584624	0.000516236473660327\\
1.22108045931707	0.00043816455573214\\
1.4393713460023	0.000374323614680307\\
1.69668579648814	0.000365840772227444\\
2	0.000365591510371301\\
2.35753726958718	0.000365513906664657\\
2.77899098874627	0.000365476494716939\\
3.27578741390813	0.000365454714661716\\
3.8613954577665	0.000365440621628377\\
4.55169185214958	0.000365430869716979\\
5.36539159055945	0.00036542380322716\\
6.32455532033676	0.000365418511164011\\
};
\addplot [color=black!15!blue, dashed, forget plot]
  table[row sep=crcr]{%
0.002	0.00036532529471884\\
0.00235753726958717	0.000365325276847856\\
0.00277899098874627	0.000365325256084637\\
0.00327578741390813	0.000365325231580991\\
0.0038613954577665	0.000365325203368745\\
0.00455169185214958	0.000365325168944493\\
0.00536539159055945	0.000365325128944818\\
0.00632455532033676	0.000365325081936541\\
0.00745518744062988	0.000365325025987641\\
0.00878794112152158	0.000365324959675264\\
0.0103589493584624	0.000365324882450253\\
0.0122108045931707	0.000365324790020126\\
0.014393713460023	0.000365324681480604\\
0.0169668579648814	0.000365324552836573\\
0.02	0.000365324400103109\\
0.0235753726958717	0.000365324220297383\\
0.0277899098874628	0.000365324006525662\\
0.0327578741390813	0.000365323753482879\\
0.038613954577665	0.000365323452097187\\
0.0455169185214958	0.000365323094446047\\
0.0536539159055945	0.00036532266797576\\
0.0632455532033676	0.000365322158475164\\
0.0745518744062988	0.000365321549566024\\
0.0878794112152158	0.000365320818149907\\
0.103589493584624	0.000365319938587151\\
0.122108045931707	0.000365318874311764\\
0.14393713460023	0.000365317582285555\\
0.169668579648814	0.000365316002996548\\
0.2	0.000365314058622255\\
0.235753726958718	0.000365311642971505\\
0.277899098874627	0.000365308609057414\\
0.327578741390813	0.000365304743408918\\
0.38613954577665	0.000365299728979115\\
0.455169185214958	0.000365293070431701\\
0.536539159055945	0.000365283948399803\\
0.632455532033676	0.000365270903144293\\
0.745518744062988	0.000365251053311899\\
0.878794112152158	0.000365217810237276\\
1.03589493584624	0.000365152096329152\\
1.22108045931707	0.000364966353060949\\
1.4393713460023	0.000362419048360648\\
1.69668579648814	0.000314582675394642\\
2	0.000267054492680019\\
2.35753726958718	0.000226600845045112\\
2.77899098874627	0.000192254148777702\\
3.27578741390813	0.000163106624307231\\
3.8613954577665	0.000138375295601023\\
4.55169185214958	0.000117392553627606\\
5.36539159055945	9.95908645393552e-05\\
6.32455532033676	8.44882821025999e-05\\
};
\addplot [color=black!50!green, dashdotted, forget plot]
  table[row sep=crcr]{%
0.002	1.53987528081038e-10\\
0.00235753726958717	1.53987528081003e-10\\
0.00277899098874627	1.53987528080962e-10\\
0.00327578741390813	1.53987528080914e-10\\
0.0038613954577665	1.53987528080857e-10\\
0.00455169185214958	1.5398752808079e-10\\
0.00536539159055945	1.5398752808071e-10\\
0.00632455532033676	1.53987528080617e-10\\
0.00745518744062988	1.53987528080507e-10\\
0.00878794112152158	1.53987528080377e-10\\
0.0103589493584624	1.53987528080224e-10\\
0.0122108045931707	1.53987528080044e-10\\
0.014393713460023	1.53987528079831e-10\\
0.0169668579648814	1.5398752807958e-10\\
0.02	1.53987528079285e-10\\
0.0235753726958717	1.53987528078937e-10\\
0.0277899098874628	1.53987528078526e-10\\
0.0327578741390813	1.53987528078042e-10\\
0.038613954577665	1.53987528077472e-10\\
0.0455169185214958	1.53987528076799e-10\\
0.0536539159055945	1.53987528076007e-10\\
0.0632455532033676	1.53987528075072e-10\\
0.0745518744062988	1.53987528073971e-10\\
0.0878794112152158	1.53987528072672e-10\\
0.103589493584624	1.53987528071142e-10\\
0.122108045931707	1.53987528069338e-10\\
0.14393713460023	1.53987528067212e-10\\
0.169668579648814	1.53987528064705e-10\\
0.2	1.53987528061751e-10\\
0.235753726958718	1.53987528058268e-10\\
0.277899098874627	1.53987528054162e-10\\
0.327578741390813	1.53987528049323e-10\\
0.38613954577665	1.53987528043618e-10\\
0.455169185214958	1.53987528036894e-10\\
0.536539159055945	1.53987528028968e-10\\
0.632455532033676	1.53987528019624e-10\\
0.745518744062988	1.53987528008611e-10\\
0.878794112152158	1.53987527995628e-10\\
1.03589493584624	1.53987527980325e-10\\
1.22108045931707	1.53987527962285e-10\\
1.4393713460023	1.53987527941021e-10\\
1.69668579648814	1.53987527915956e-10\\
2	1.53987527886409e-10\\
2.35753726958718	1.53987527851581e-10\\
2.77899098874627	1.53987527810526e-10\\
3.27578741390813	1.53987527762132e-10\\
3.8613954577665	1.53987527705087e-10\\
4.55169185214958	1.53987527637843e-10\\
5.36539159055945	1.53987527558579e-10\\
6.32455532033676	1.53987527465145e-10\\
};
\end{axis}
\end{tikzpicture}%
}

  \newsavebox\insetplot
  \savebox\insetplot{%
  \scriptsize

\begin{tikzpicture}[
    scale = 0.3,
    box/.style = {rectangle, draw=black, fill=white, thick, minimum width=2.5cm, text width=2.5cm, align=center},
  ]
  \node[box] (qp) at (0, 0) {superconductor quasiparticles ($T_{qp}$)};
  \node[box] (ph) [below=0.5cm of qp] {superconductor phonons ($T_{ph}$)};
  \node[box] (sub) [below=0.5cm of ph] {substrate phonons ($T_{su}$)};
  \node[box] (bath) [below=0.5cm of sub] {bath ($T_b$)};
  \draw[->] (qp) to node[auto] {$G\tu{I} \!\sim\! \si{pW/K}$} (ph);
  \draw[->] (ph) to node[auto] {$G\tu{P} \!\sim\! \si{nW/K}$} (sub);
  \draw[->] (sub) to node[auto] {$G\tu{L} \!\sim\! \si{pW/K} \!-\! \si{nW/K}$} (bath);

  \node[right=0.1cm of qp] {$C\tu{qp} \!\sim\! \si{aJ/K}$};
  \node[right=0.1cm of ph] {$C\tu{ph} \!\sim\! \si{aJ/K}$};
  \node[right=0.1cm of sub] {$C\tu{su} \!\sim\! \si{fJ/K}$};

\end{tikzpicture}

  }
  \tikzexternalenable
  \fi

  \tikzsetnextfilename{el_figure_1_time_constants}
  \begin{tikzpicture}
    \node[anchor=south west, inner sep=0] (myplot) at (0,0) {\usebox\mainplot};
    \begin{scope}[x={(myplot.south east)}, y={(myplot.north west)}]
      \node[anchor=south west, inner sep=0] at (0.225,0.175) {\usebox\insetplot};
    \end{scope}
  \end{tikzpicture}

  \caption{Variation of the dynamical time constants (from the eigenvalues of \cref{eq:eigenproblem} in the frequency domain) as the device thermal isolation from the bath is decreased by changing the geometry factor $g_{su}$. With $P_R = \SI{25}{fW}$, and zero readout frequency detuning ($\nu_r = \nu_0(T_{qp})$). Inset: Block diagram of the thermal model of a generic Kinetic Inductance Detector with thermal isolation. Three heat capacities, each with their own effective temperatures $T_i$, and relevant power flows $q_i$ are considered. Typical heat capacities $C$ and thermal conductances $G$ for the device design and operating conditions described in the text are labelled.}
  \label{fig:1}
\end{figure}
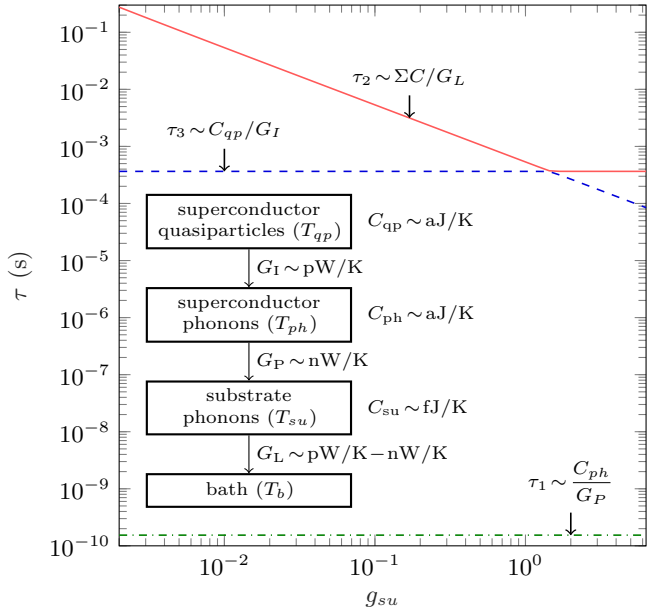

These expressions for $q_I$ and $q_R$ are plotted against quasiparticle temperature $T_{qp}$ in \cref{fig:5} for an example resonator. Considering only $T_{qp}$, possible steady state operating points where there is no net power flowing into the quasiparticles are at the intersections of $q_I$ and $q_R$. We note that for negative detuning of the readout frequency ($\Delta\nu = \nu_r - \nu_0 < 0$), in this example there are three possible operating points -- two stable and one unstable -- indicated by the arrows. On a frequency sweep, this shows as hysteresis with respect to sweep direction of the quasiparticle temperature~\cite{Thompson2013} and hence the scattering parameter $\abs{S_{21}}$. We also contrast the superconducting two-temperature exponential model for $q_I$ \cref{eq:super_q} to a normal state power law electron-phonon expression for metals
\begin{equation}
q_I\ts{normal} = V \Sigma_n (T_{qp}^5 - T_{ph}^5) \; , \label{eq:normal_q}
\end{equation}
with $\Sigma_n = \SI{1.68E9}{W.m^{-3}.K^{-5}}$ for Al from the material properties used. At low temperatures, the normal state power flow is several orders of magnitude greater than the superconducting value from \cref{eq:super_q}. If implemented in calculations, this would reduce or hide the effects of readout power heating.

For the power flow between the superconductor phonons and the substrate phonons we use the Kapitza expression $q_P = \Sigma_{su} A_{res}(T_{ph}^4 - T_{su}^4)$, where $A_{res}$ is the area of the superconducting resonator in contact with the substrate and $\Sigma_{su}$ is a material-dependent constant, $\SI{850}{W.m^{-2}.K^{-4}}$ for an Al/\SiN{} interface~\cite{DeVisser2010}. The power flow from the substrate to bath is modelled as effectively limited by narrow legs. The thermal power carried by \SiN{} legs is found empirically~\cite{Rostem2008} to be $q_L = g_{su} K_{b}(T_{su}^n - T_{b}^n)$, where $g_{su}$ is a geometric factor equal to the ratio of the width to length of the legs, $K_b$ is a material-dependent constant, and the exponent $n$ depends on the detailed physics of the heat flow. Experiments in our group for \SI{200}{nm} thick \SiN{} find, typically, $K_b = \SI{60}{pW.K^{\mathit{-n}}}$ and $n = 2$~\cite{Goldie2011}.

Including both the signal power $q_S = \eta_S P_S$ absorbed by the quasiparticles, and heater power $q_H = \eta_H P_H$ absorbed by the substrate phonons, allows consideration of both a direct photon detection mode of operation, and a bolometric mode in which the KID resonator is thermally coupled to some other absorbing system. In this work, we assume the absorbed power is simply the incident power ($P_S$ and $P_H$) multiplied by an efficiency ($\eta_S$ and $\eta_H$) but in general more complex models for absorbed power could be incorporated.

Following \refcite{Thomas2014}, we combine the elements into a temperature vector $\vc{T}$, a heat capacity matrix $\mat{C_0}(\vc{T}, \vc{v})$ and a power flow vector $\vc{q}(\vc{T}, \vc{v})$ which describes the net power flow into each element of the model. The vector $\vc{v}$ includes all other parameters which do not depend on temperature, such as readout frequency $\nu_r$, readout applied power $P_R$, signal frequency $\nu_s$, signal power $P_S$, and power into the substrate $P_H$. Finding the steady-state operating point of the device involves finding the temperature vector $\vc{T_0}$ such that $\vc{q}(\vc{T_0}, \vc{v_0}) = \vc{0}$, for a given parameter vector $\vc{v_0}$.
The effective thermal conductance matrix $\mat{G_0} = \eval{-\pdv*{\vc{q}}{\vc{T}}}_{\vc{v_0}}$ then includes all contributions to the dynamical behaviour of the device, including electrothermal feedback.
Typical values for the heat capacities $C$ and thermal conductances $G$ are provided in the block diagram inset of \cref{fig:1}. The substrate-bath conductance $G_L$ is varied by changing the geometry factor $g_{su}$ over a few orders of magnitude.

The dynamical behaviour of the device about the chosen operating point is described by
\begin{equation}
  \pd{\Delta\vc{T(t)}}{t} = -\mat{C_0}^{-1} \vdot \mat{G_0} \vdot \Delta\vc{T}(t) \;, \label{eq:eigenproblem}
\end{equation}
which can be analysed in the frequency domain as an eigenvalue problem. The modes and dynamical time constants are found by diagonalising $\mat{G_0}^{-1} \vdot \mat{C_0}$. As $\mat{G_0}$ is not diagonal, the effective time constants may be quite different from the time constants expected from $C/G$ of any individual elements. There are three dynamical modes: in the regime where the limiting conductance is the quasiparticle-phonon cooling $q_I$, the modes are a fast cooling of the superconductor phonons into the substrate phonons; a moderately slow cooling of the entire device into the bath; and a slow cooling of the superconductor quasiparticles into the superconductor phonons. In general in this regime we find $T_{qp} > T_{ph} \approx T_{su} \approx T_{b}$.
\Cref{fig:1} shows how the three time constants associated with these change as the device's thermal isolation from the bath is increased, by changing the geometric factor $g_{su}$ associated with the substrate-bath conductance $G_L$. As $G_L$ is reduced, the dynamical modes change so that the overall device cooling is slower than the superconductor quasiparticle to phonon cooling, and now we find $T_{ph} > T_{b}$ and $T_{su} > T_{b}$. In the figure, the time constants are labelled based on the heat capacities and conductances involved in the modes they are associated with; for example, $\tau_1$, the smallest time constant, is the eigenvalue for the fastest dynamical mode involving power flowing from the superconductor phonons into the substrate phonons, as for typical parameters this involves a large conductance $G_P$ compared to the quasiparticle-phonon conductance $G_I$~\cite{DeVisser2010}.

\subsection{Numerical implementation}

The software implementation of the model contains two parts: finding the operating point given a set of parameters (``large-signal''), and linearising the behaviour of the device in response to small perturbations about that operating point (``small-signal'').

Low temperature ($T \ll T_c$, the superconducting critical temperature) analytical approximations for superconductor complex conductivity~\cite{Tinkham2004} and Debye phonon heat capacity were used.
Superconductor quasiparticle heat capacity and superconducting energy gap were calculated by numerical integration, assuming the quasiparticle energy distribution could be approximated by Fermi functions at the quasiparticle effective temperature $T_{qp}$~\cite{Tinkham2004}.
Solving for the operating point $\vc{q}(\vc{T}, \vc{v}) = \vc{0}$ for $\vc{T} = \begin{bmatrix} T_{qp} & T_{ph} & T_{su} \end{bmatrix}$ given $\vc{v} = \begin{bmatrix}P_R & \nu_r & P_S & \nu_s & P_H \end{bmatrix}$ was done by multidimensional Newton-Raphson iteration with a quadratic line search~\cite{Kelley1995}, and with all temperatures bounded between $T_b$ and $T_c$. Analytical expressions for derivatives were used, except for $\pdv*{S_{21}}{T}$ and $\pdv*{S_{21}}{T}$ which were approximated numerically using first-order central differences.
In some cases, we want the operating point to follow the resonant frequency even as it changed with quasiparticle temperature or other parameters. However, the steady state operating point is needed to calculate the resonant frequency, and changing the readout frequency changes the operating point. Therefore these calculations, described as at fixed detuning rather than at a fixed readout frequency, were done by iteratively solving for the operating point and appropriate readout frequency, until the relative readout frequency change in one iteration was less than \num{E-9}.

\subsection{Modelled device parameters}
In this work, we consider a device configuration in which an Al superconducting resonator is patterned on a \SiN{} membrane, which in turn is suspended by thin \SiN{} legs from the surrounding bulk substrate.
The device consists of a resonator of dimensions $\SI{4}{mm} \times \SI{3}{\micro{}m} \times \SI{200}{nm}$, patterned on a \SiN{} membrane of the same dimensions, connected effectively by a single thin \SiN{} leg of dimensions $\SI{2}{\micro{}m} \times \SI{100}{\micro{}m} \times \SI{200}{nm}$ to the thermal bath, which is always at $T_b = 0.1\,T_c = \SI{0.118}{K}$.
The superconducting Al was set to have a zero-temperature energy gap of $\Delta_0 = \SI{180}{\micro{}eV}$, critical temperature $T_c = \SI{1.18}{K}$, and residual resistivity $\rho_n(\SI{0}{K}) = \SI{5.8E-9}{\ohm.m}$, all typical of films deposited by our group~\cite{Vardulakis2008}.

At a bath temperature $T_b = 0.1\,T_c$, and a resonance frequency of $\nu_0 = \SI{5}{GHz}$, we require $Q_I \ge \num{E8}$ from \cref{eq:qi_condition}. In all the simulations reported here we choose $Q_I(T_b) = \num{6E8}$ and $Q_C(T_b) = \num{3E5}$, with a generator/load impedance $Z_0 = \SI{50}{\ohm}$. These are the values of $Q_I$ and $Q_C$ at zero incident power; at a typical operating point, when $T_{qp} > T_b$, $Q_I$ is much reduced. %

We assume the incident signal power $P_S$ is dissipated in the quasiparticles with 100\% efficiency ($\eta_S = 1$), independent of signal frequency, as with the incident power $P_H$ on the substrate ($\eta_H = 1$). We also ignore any cable loss or change to the signal from the mixer or filtering ($R_I, R_Q, R_0 = 1$ from \refcite{Thomas2014}).

\section{Results} \label{sec:results}
\subsection{Operating point}

\begin{figure}[htb]
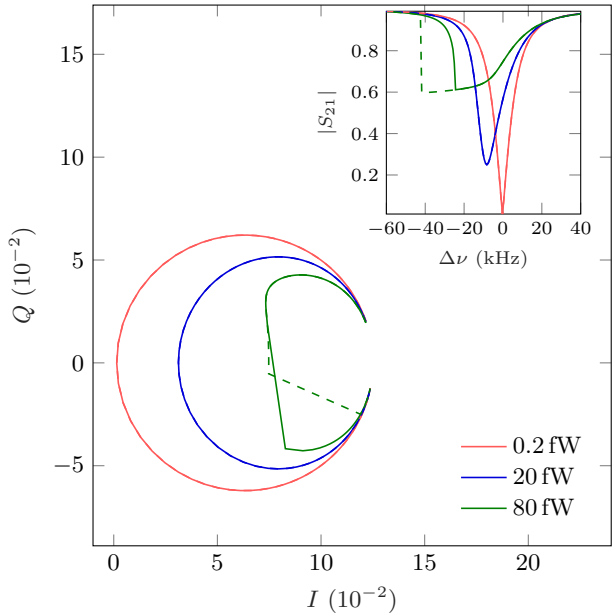

  \centering

  \ifhavepgf
  \tikzset{external/remake next}
  \tikzexternaldisable
  \savebox\mainplot{\input{figures/el_figure_4_IQ_power.tikz}}

  \savebox\insetplot{%
    \scriptsize
    \setlength{\figureheight}{0.15\textwidth}
    \setlength{\figurewidth}{0.15\textwidth}
    \input{figures/el_figure_3_resonance_power.tikz}
  }
  \tikzexternalenable
  \fi

  \tikzsetnextfilename{el_figure_4_IQ_power}
  \begin{tikzpicture}
    \node[anchor=south west, inner sep=0] (myplot) at (0,0) {\usebox\mainplot};
    \begin{scope}[x={(myplot.south east)}, y={(myplot.north west)}]
      \node[anchor=north east, inner sep=0] at (0.99,0.99) {\usebox\insetplot};
    \end{scope}
  \end{tikzpicture}

  \caption{$IQ$ mixer outputs during up (solid) and down (dashed) sweeps of readout frequency, at selected incident readout powers $P_R$. At high readout powers, hysteretic behaviour is observed -- the up and down frequency sweeps follow different trajectories in the $IQ$ plane. Inset: The forward scattering parameter $\abs{S_{21}}$ measured at the device plane for the same frequency sweeps.}
  \label{fig:4}
\end{figure}

By sequentially solving for the operating point, we calculate how the operating point changes as we sweep the readout frequency, both in the forward and reverse directions. The inset of \cref{fig:4} plots the scattering parameter $\abs{S_{21}}$ for forward and reverse frequency sweeps at three different incident readout powers. At the highest readout power plotted, the forward (solid line) and reverse (dashed line) do not follow the same path -- readout power heating induced hysteresis~\cite{Thompson2013}.
The exact readout power at which hysteresis appears is dependent on the resonator volume, as well as the efficiency with which the resonator couples to the readout line, and the bath temperature.
At optimal coupling $Q_I = Q_C$, and for low bath temperatures, we expect larger effects due to heating, and so visible nonlinearities at lower powers.

In practice, an $IQ$ readout is usually used for reading out the device, which allows measurement of both the amplitude and phase of the readout tone as the in-phase and quadrature outputs. The main plot of \cref{fig:4} shows the same frequency sweep as the inset, now plotted in the $IQ$ plane. For a perfectly symmetric resonance, a frequency sweep traces out a circle. As the readout power increases, increasing the quasiparticle temperature, the circle radius decreases; at the highest powers, the circle distorts and shows hysteresis between the readout frequency up (solid line) and down (dashed line) sweeps.

This phenomena is similar in effect, but different in source, to hysteresis due to the intrinsic nonlinearity of the kinetic inductance~\cite{Zmuidzinas2012,Swenson2013}, which is not included in this work, as the emphasis here is on electrothermal feedback due to readout power heating.
In particular, readout power heating causes a dissipative nonlinearity which causes both a resonant frequency shift and a significant change in the depth of the resonance~\cite{Thompson2013}, whereas a purely reactive nonlinearity such as the kinetic inductance nonlinearity causes primarily a shift in resonance frequency~\cite{Swenson2013}.
Microscopically, the heating effect corresponds to the heating of quasiparticles by the microwave field, while the intrinsic kinetic inductance nonlinearity corresponds to the changes in the Cooper pair states due to the microwave field~\cite{Semenov2016}. Which phenomenon is more significant depends on the detailed geometry of the device, as the heating effect is a function of the absorbed power density, while the kinetic inductance nonlinearity is a function of the current density.

\begin{figure}[htb]
  \centering
  \tikzsetnextfilename{el_figure_6_operating_point}
  \ifhavepgf
  \pgfplotsset{legend style = {font=\scriptsize}}
  \fi
  \input{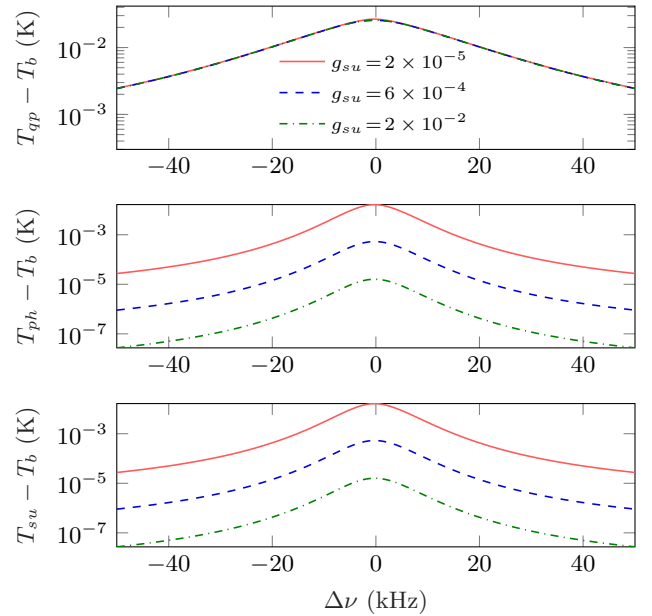}
  \caption{Steady-state temperatures of superconductor quasiparticles ($T_{qp}$), superconductor phonons ($T_{ph}$), and substrate phonons ($T_{su}$) as a function of readout frequency detuning ($\Delta\nu = \nu_r - \nu_0$), at three different values for the thermal conductance of the device to the bath (by changing $g_{su}$). Maximum heating is at zero detuning. Increasing the thermal isolation of the device increases $T_{ph}$ and $T_{su}$ but not $T_{qp}$. With $P_R = \SI{0.2}{fW}$.}
  \label{fig:6}
\end{figure}

Typically, the limiting thermal conductance is the quasiparticle to phonon conductance $G_I$. The steady-state operating point therefore usually has $T_{qp} > T_{ph} \approx T_{su} \approx T_{b}$. We plot the operating point temperatures in \cref{fig:6} as a function of readout frequency detuning; the largest readout power absorbed, and consequently heating, is at zero detuning. If we increase the thermal isolation of the device from the bath by decreasing $g_{su}$ and hence $G_L$, the operating point shifts so both $T_{ph} > T_{b}$ and $T_{su} > T_{b}$, but still $T_{ph} \approx T_{su}$; interestingly, however, $T_{qp}$ does not change by the same magnitude -- when there is significant quasiparticle heating, $T_{qp}$ is isolated from changes to the other temperatures. This is due to the functional form of $G_I$ being highly nonlinear even for moderate absorbed powers; if we instead use the normal-state power law expression \cref{eq:normal_q}, $T_{qp}$ changes by the same amount as $T_{ph}$ and $T_{su}$, as in a linear system where $G$ is constant. This effect may be useful for isolating the responsivity of the device from small changes in bath temperature, a potential source of noise.

\subsection{Small-signal analysis}

Once the operating point of the device is fixed, we can consider the response to small perturbations.
Linearising the $IQ$-response about the operating point for small time-dependent changes in the parameter vector $\Delta \vc{v}(f)$, we find
  \begin{align}
    \Delta{}I(f) &= F_I(f) \; \hat{\vc{x}} \vdot \mat{K_0}(f) \vdot \Delta\vc{v}(f) + H_I(f) \frac{\Delta{}P_R(f)}{2P_R}\\
    \Delta{}Q(f) &= F_Q(f) \; \hat{\vc{x}} \vdot \mat{K_0}(f) \vdot \Delta\vc{v}(f) + H_Q(f) \frac{\Delta{}P_R(f)}{2P_R} \;,
  \end{align}
in the frequency domain. $F_{I,Q}(f)$ describes the $I$ and $Q$ response to changes in quasiparticle temperature; $\mat{K_0}(f)$ maps changes in the parameter vector $\vc{v}$ to changes in the temperature vector $\vc{T}$; and $H_{I,Q}(f)$ describes the change in $I$ and $Q$ directly due to changes in readout power, independent of quasiparticle temperature. A full derivation and discussion is available in \refcite{Thomas2014}.

Full expressions for the readout contributions to $\mat{K_0}$ in \textcite{Thomas2014} account for the resonator dynamics -- in particular, its electrical response time $Q_T/\nu_0$. This introduces imaginary components in the frequency domain, and physically corresponds to the possibility of energy oscillating between the electrical and thermal systems.
All other contributions to $\mat{K_0}$ are assumed quasistatic, and so are entirely real.

\begin{figure}[htb]
  \centering
  \tikzsetnextfilename{el_figure_7_effective_conductance}
  \input{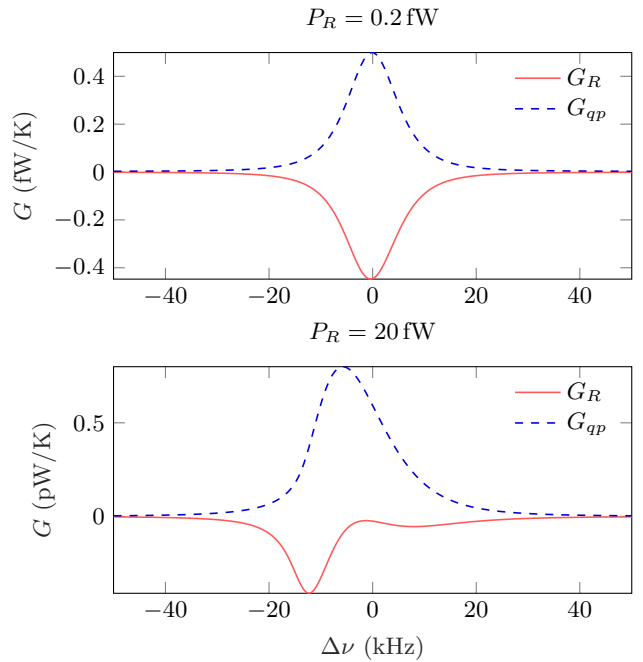}
  \caption{(a): Effective thermal conductance between superconductor quasiparticles and bath as a function of readout frequency detuning $\Delta\nu = \nu_r - \nu_0$. $G_R = -\pdv*{q_R}{T_{qp}}$ (solid) is the contribution of electrothermal feedback to the effective thermal conductance, and $G_{qp} = -\pdv*{q_R}{T_{qp}} + \pdv*{q_I}{T_{qp}}$ (dashed) is the total component of thermal conductance dependent on quasiparticle temperature. (b): At high readout powers, $T_{qp}$ asymmetry about the resonant frequency also leads to an asymmetric $G_R$.}
  \label{fig:7}
\end{figure}

We identify the contribution to $\mat{G_0}$ from $G_R = -\pdv*{q_R}{T_{qp}}$ as the electrothermal feedback. As shown in \cref{fig:7}, $G_R$ has the opposite sign to the other contributions to the quasiparticle temperature-dependent elements of the thermal conductance matrix $\qty{\mat{G_0}}_{1,1} = G_{qp}$, and so has the effect of reducing the net effective thermal conductance from the quasiparticles to the superconductor phonons. In principle, it may be possible to find a device operating point where the contribution of electrothermal feedback is made into negative feedback, and so enhances the net effective thermal conductance, by varying the readout frequency, device quality factors and readout powers. However, stable device operation is only possible for a limited range of parameters, and we find for a typical configuration the contribution of electrothermal feedback is consistently positive. At high readout power, when the operating point (in particular $T_{qp}$) is asymmetric about the resonance frequency due to readout power heating, $G_R$ is also asymmetric about the resonance frequency. %

\begin{figure}[htb]
  \centering
  \setlength{\figureheight}{0.38\textwidth}
  \setlength{\figurewidth}{0.35\textwidth}
  \tikzsetnextfilename{el_figure_8_thermal_electrical_response}
  \input{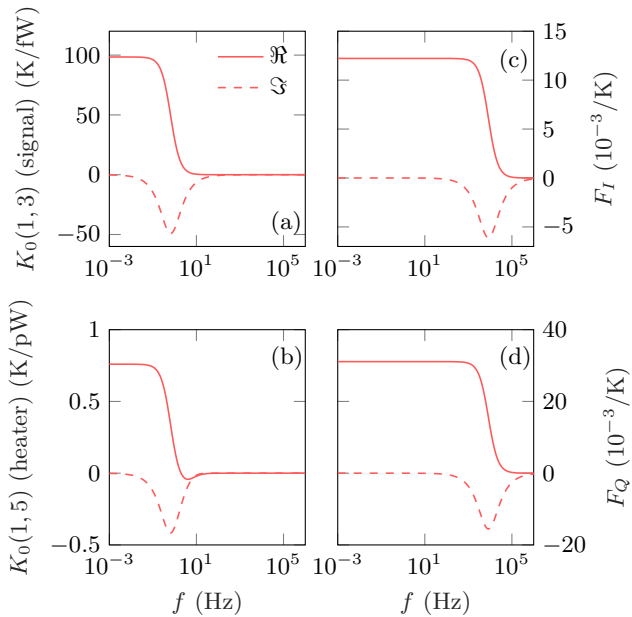}
  \caption{(a) and (b): Real (solid) and imaginary (dashed) parts of elements of the parameter response matrix $\mat{K_0}$, which maps changes in the parameter vector $\vc{v}$ to changes in the temperature vector $\vc{T}$, as a function of the modulation frequency of $\Delta\vc{v}(f)$. The response is bandwidth limited by the electrothermal dynamical response time of the overall device. $\qty{K_0}_{1,3}$ is the superconductor quasiparticle temperature $T_{qp}$ response to changes in absorbed signal power $P_S$, and $\qty{K_0}_{1,5}$ is the $T_{qp}$ response to changes in absorbed substrate heater power $P_H$.
  (c) and (d): Real (solid) and imaginary (dashed) parts of $IQ$ output response $F_{I,Q}$ which map changes in superconductor quasiparticle temperature $T_{qp}$ to changes in $I$ and $Q$ mixer outputs, as a function of the modulation frequency of $\Delta\vc{v}(f)$. This response is limited by the electrical bandwidth of the resonator, $\nu_0/Q_T$.
  With zero detuning of readout frequency, and $P_R = \SI{20}{aW}$.}
  \label{fig:8}
\end{figure}

$\mat{K_0}(f)$ maps small changes in the parameter vector $\vc{v}$ to changes in the temperature vector $\vc{T}$. Its bandwidth is determined by the time constants associated with the dynamical cooling modes of the device. We can compare the magnitude of the response to changes in signal power to changes in substrate heating power to understand the operation of the device as a phonon-mediated detector, where a separate structure absorbs the primary signal and is connected to the device substrate only thermally, and so substrate phonons must carry power to the KID for detection~\cite{Timofeev2014,Cardani2015}. \Cref{fig:8} shows the photon signal power response in (a) is several orders of magnitude greater than the substrate heating power response in (b). This would again appear to be due to the nonlinearity of the quasiparticle-phonon conductance effectively isolating the quasiparticle temperature from the rest of the device
-- if using the normal-state power law expression \cref{eq:normal_q} the responsivities are equal, but when using the appropriate superconducting expression \cref{eq:super_q}, orders of magnitude more substrate heating power is required than pair-breaking photon power in order to cause the same change in quasiparticle temperature.
$F_{I,Q}$ is the response of the $IQ$ output to changes in quasiparticle temperature $T_{qp}$. The rolloff of the electrical response outside the bandwidth of the detector ($\nu_0/Q_T$) is seen in \cref{fig:8}(c) and (d). This rolloff of the electrical response is also dependent on readout frequency detuning.

\subsection{Responsivity}

To quantify the responsivity of the device to signal power $P_S$, we can investigate the expressions $\Delta{}I/\Delta{}P_S$ and $\Delta{}Q/\Delta{}P_S$, where $\Delta{}I$ and $\Delta{}Q$ are the changes in the $I$ and $Q$ outputs for a change in signal power $\Delta{}P_S$. We can consider the resulting trajectory in the $IQ$ plane, combining these two expressions to obtain a direction of maximum responsivity.
A single frequency modulation in signal power creates an ellipse centred on the operating point. The major and minor axes of this ellipse are
  \begin{align}
    \Delta{}A(f) &= \sqrt{\Re{\Delta{}I(f)}^2 + \Re{\Delta{}Q(f)}^2} \\
    \Delta{}B(f) &= \sqrt{\Im{\Delta{}I(f)}^2 + \Im{\Delta{}Q(f)}^2} \;. \label{eq:AB}
  \end{align}
By choosing the maximum of $\Delta{}A/\Delta{}P_S$ and $\Delta{}B/\Delta{}P_S$ we have a single measure of the responsivity of the device when using an optimal readout technique.

\begin{figure}[htb]
  \centering
  \tikzsetnextfilename{el_figure_17_responsivity_detuning_feedback}
%
%
\begin{tikzpicture}

\begin{axis}[%
width=0.951\figurewidth,
height=\figureheight,
at={(0\figurewidth,0\figureheight)},
scale only axis,
xmin=-52.3114525680542,
xmax=50.8829405050278,
xlabel style={font=\color{white!15!black}},
xlabel={$\Delta\nu$ (\si{kHz})},
ymin=0,
ymax=6,
ylabel style={font=\color{white!15!black}},
ylabel={$\Delta{}A/\Delta{}P_S$ (1/pW)},
legend style={legend cell align=left, align=left, fill=none, draw=none}
]
\addplot [color=white!35!red]
  table[row sep=crcr]{%
-52.3114525680542	3.39145229438558\\
-51.3469808359146	3.39709554614451\\
-50.3807222776413	3.40221666362413\\
-49.4162190666199	3.4179342785781\\
-48.4535711669922	3.42218438721145\\
-47.4928868103027	3.42603478801905\\
-46.5342916097641	3.440453663477\\
-45.5779147033691	3.44345944546373\\
-44.6238974866867	3.45704972750217\\
-43.6723876218796	3.47025122289855\\
-42.7235590524673	3.47208125168246\\
-41.777586022377	3.48453122662518\\
-40.8346609649658	3.49666243071246\\
-39.894994099617	3.50846648359943\\
-38.9588148746491	3.53104359227838\\
-38.0263783121109	3.54236817877037\\
-37.0979399347305	3.55352025521839\\
-36.1738061542511	3.57565299217569\\
-35.2595088214874	3.59801297024392\\
-34.3459388046265	3.62033546471078\\
-33.4379286251068	3.64288172208081\\
-32.5359467763901	3.6657972349658\\
-31.6405044517517	3.68922451483897\\
-30.752189332962	3.72488252589116\\
-29.8716252622604	3.74998184671996\\
-28.9942606210709	3.78779455509347\\
-28.1297122573853	3.83941546971535\\
-27.2746973762512	3.88122229158201\\
-26.4299482326508	3.92545625571545\\
-25.6025496921539	3.9853983765276\\
-24.7838314266205	4.0491030702339\\
-23.9731747093201	4.13053438136678\\
-23.1803024358749	4.20406642547085\\
-22.411022187233	4.29771218877797\\
-21.65129641819	4.3980697825531\\
-20.9060832567215	4.52102223353472\\
-20.1895220499039	4.63509582762313\\
-19.4874004745483	4.78659967508462\\
-18.7997560386658	4.92408954044822\\
-18.1398224277496	5.05204741660881\\
-17.4657120494843	5.19326959313531\\
-16.5895724411011	5.26272000346216\\
-15.7650143117905	5.18775028754118\\
-15.4828045911789	5.10890473973269\\
-14.8937410945892	4.87140111598952\\
-14.1720927343369	4.48478097442084\\
-13.3692919721603	3.99328801664671\\
-12.4544202337265	3.49978100853046\\
-11.399892580986	3.03485541845676\\
-10.1734951105118	2.65433297445424\\
-8.78393710231781	2.36215478710805\\
-7.24736009311676	2.15578260332394\\
-5.59389046287537	2.0295324956822\\
-3.87399905014038	1.9605067381016\\
-2.12698771286011	1.94254870001991\\
-0.38716445350647	1.95624214494287\\
1.32060879135132	1.99660920239235\\
2.98084754943848	2.05209405376316\\
4.58578451824188	2.11889443894772\\
6.13324117469788	2.19185618703012\\
7.62464794826508	2.26702845351881\\
9.05974321842194	2.34505946199415\\
10.4517645683289	2.42099806819418\\
11.8001243066788	2.49125174752043\\
13.1096071243286	2.55939327497291\\
14.3847393512726	2.61998085103377\\
15.6247146606445	2.67906427486398\\
16.8446795415878	2.73255025610403\\
18.0408227424622	2.78564313636415\\
19.2160905199051	2.82692170134349\\
20.3730422906876	2.86847519480208\\
21.5139253702164	2.90460310800693\\
22.6406734962463	2.94230697077192\\
23.7500669727325	2.97511270747354\\
24.8589732570648	2.99702863461754\\
25.9519212436676	3.02469905721309\\
27.0369163560867	3.04942396821789\\
28.1142680978775	3.06632968976414\\
29.1848167228699	3.0875993099897\\
30.2493008918762	3.10582321201205\\
31.3083624591827	3.12117588953553\\
32.3625698251724	3.13380671972867\\
33.4124269924164	3.1438598139861\\
34.4583815326691	3.15147420420926\\
35.5008159513473	3.16546550466906\\
36.5400888996124	3.17750798353997\\
37.5764953927994	3.1787417607994\\
38.6103191766739	3.18702628126309\\
39.6418048448563	3.19360986612934\\
40.6711712236404	3.19861135370785\\
41.6986152877808	3.21137993476942\\
42.7243046731949	3.21345130766622\\
43.7484074697495	3.21411967144975\\
44.7710630598068	3.22302516641766\\
45.7923964252472	3.22114463294601\\
46.812525229454	3.22776094213673\\
47.8315566196442	3.23333451131624\\
48.8495794706345	3.22806346419471\\
49.8666808643341	3.23160911468905\\
50.8829405050278	3.23423642426783\\
};
\addlegendentry{with feedback}

\addplot [color=black!15!blue, dashed]
  table[row sep=crcr]{%
-52.3114525680542	1.80270390288014\\
-51.3469808359146	1.80416522865272\\
-50.3807222776413	1.80523786041159\\
-49.4162190666199	1.81160563491122\\
-48.4535711669922	1.81191955512785\\
-47.4928868103027	1.81186337069609\\
-46.5342916097641	1.81701790255024\\
-45.5779147033691	1.81615419326686\\
-44.6238974866867	1.8204288026702\\
-43.6723876218796	1.82423738662684\\
-42.7235590524673	1.82205439112564\\
-41.777586022377	1.8249039222353\\
-40.8346609649658	1.827243035325\\
-39.894994099617	1.82905616485854\\
-38.9588148746491	1.83572368034055\\
-38.0263783121109	1.83640080333613\\
-37.0979399347305	1.83650486280008\\
-36.1738061542511	1.84132362927461\\
-35.2595088214874	1.84560505378463\\
-34.3459388046265	1.84908616530444\\
-33.4379286251068	1.85186070306575\\
-32.5359467763901	1.85389976845481\\
-31.6405044517517	1.85518842380088\\
-30.752189332962	1.86083162175532\\
-29.8716252622604	1.86051428574734\\
-28.9942606210709	1.86428088992806\\
-28.1297122573853	1.87227715729375\\
-27.2746973762512	1.87424167420117\\
-26.4299482326508	1.87522426848908\\
-25.6025496921539	1.8803466763206\\
-24.7838314266205	1.88424441780873\\
-23.9731747093201	1.8916729721446\\
-23.1803024358749	1.89318258199247\\
-22.411022187233	1.89866766927136\\
-21.65129641819	1.90262360352201\\
-20.9060832567215	1.90989821292733\\
-20.1895220499039	1.91174670821827\\
-19.4874004745483	1.92094774286992\\
-18.7997560386658	1.92520496330609\\
-18.1398224277496	1.92855213366195\\
-17.4657120494843	1.93842930981969\\
-16.5895724411011	1.94396438036219\\
-15.7650143117905	1.95338768092727\\
-15.4828045911789	1.95544182981508\\
-14.8937410945892	1.9604876611142\\
-14.1720927343369	1.96870956804135\\
-13.3692919721603	1.97314247347937\\
-12.4544202337265	1.98384899342007\\
-11.399892580986	1.98876707387529\\
-10.1734951105118	1.99584100974012\\
-8.78393710231781	1.99584243960271\\
-7.24736009311676	1.98873264004074\\
-5.59389046287537	1.98386261665597\\
-3.87399905014038	1.97316727833396\\
-2.12698771286011	1.96874198003075\\
-0.38716445350647	1.96058863315997\\
1.32060879135132	1.95567694704956\\
2.98084754943848	1.94828638098303\\
4.58578451824188	1.94172838883026\\
6.13324117469788	1.93526990095276\\
7.62464794826508	1.92840550266145\\
9.05974321842194	1.92524118914988\\
10.4517645683289	1.9212958851939\\
11.8001243066788	1.91478254117657\\
13.1096071243286	1.91002601385814\\
14.3847393512726	1.90254343366064\\
15.6247146606445	1.89841709966943\\
16.8446795415878	1.89325225085764\\
18.0408227424622	1.8917138324401\\
19.2160905199051	1.88414043748638\\
20.3730422906876	1.88029443508783\\
21.5139253702164	1.87535755697411\\
22.6406734962463	1.87434426368279\\
23.7500669727325	1.8722334290749\\
24.8589732570648	1.86436268805817\\
25.9519212436676	1.86268425965754\\
27.0369163560867	1.86079383243413\\
28.1142680978775	1.85516163698373\\
29.1848167228699	1.85388829205715\\
30.2493008918762	1.85185717110431\\
31.3083624591827	1.84909298379961\\
32.3625698251724	1.84561952026575\\
33.4124269924164	1.84144992725305\\
34.4583815326691	1.83660822189914\\
35.5008159513473	1.83649463955791\\
36.5400888996124	1.83580521211301\\
37.5764953927994	1.82913105736232\\
38.6103191766739	1.82731480463325\\
39.6418048448563	1.82497168512644\\
40.6711712236404	1.82211890020771\\
41.6986152877808	1.82429681258397\\
42.7243046731949	1.82048596150787\\
43.7484074697495	1.81620649716782\\
44.7710630598068	1.81706694485393\\
45.7923964252472	1.81192255189488\\
46.812525229454	1.81197602236349\\
47.8315566196442	1.81165750199507\\
48.8495794706345	1.80529188530993\\
49.8666808643341	1.80422357769155\\
50.8829405050278	1.80281471759121\\
};
\addlegendentry{without feedback}

\end{axis}
\end{tikzpicture}%
  \caption{Responsivity of $IQ$ output, in the direction of maximum responsivity $A$, to zero-frequency changes in signal power $P_S$, with (solid) and without (dashed) the effects of electrothermal feedback, as a function of readout frequency detuning $\Delta\nu = \nu_r - \nu_0$. At this moderately high incident readout power, heating leads to an operating point asymmetric about resonance, and leads to the peak responsivity being slightly below resonance. Including the contribution of electrothermal feedback decreases the effective thermal conductance, increasing the responsivity.
  With $P_R = \SI{25}{fW}$.}
  \label{fig:17}
\end{figure}
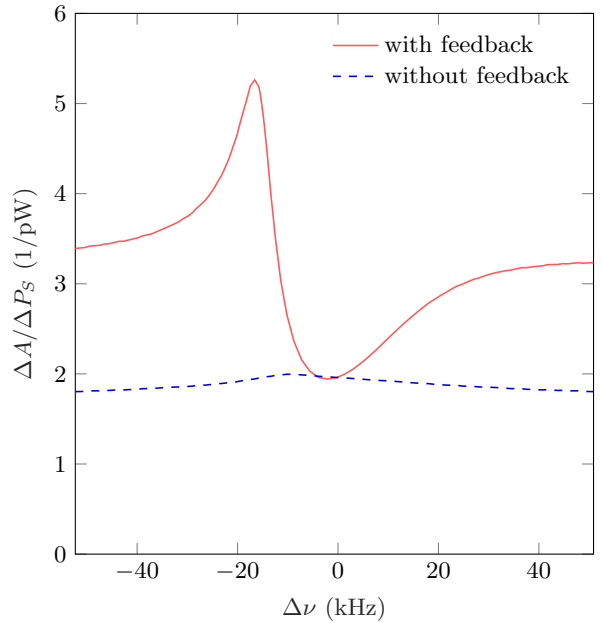

One of the primary aims of this work is to extract from the model the dependence of the responsivity on electrothermal feedback and the thermal isolation of the device. If we construct an effective overall thermal conductance between the quasiparticles and the bath $G\tu{total}$, which must always be positive at a stable operating point, we would expect the responsivity essentially depends on the inverse of this effective thermal conductance, $\propto G\tu{total}^{-1}$, as in standard bolometer theory~\cite{Timofeev2014}. Therefore the effect of feedback -- under typical conditions, a negative contribution reducing the positive $G\tu{total}$ -- is to increase responsivity, and this is confirmed by the detailed model as shown in~\cref{fig:17}.
For a symmetric resonance, the maximum responsivity is obtained at zero detuning, when the readout frequency $\nu_r$ is equal to the resonant frequency $\nu_0(T_{qp})$. When the resonance is asymmetric due to readout power heating, the optimal readout frequency may be below the resonant frequency, also shown in~\cref{fig:17}. Readout schemes exploiting nonlinear, asymmetric resonances may be possible for devices operating within a carefully controlled parameter space~\cite{Swenson2013}.

\begin{figure}[htb]
  \centering
  \tikzsetnextfilename{el_figure_15_responsivity_isolation}
  \input{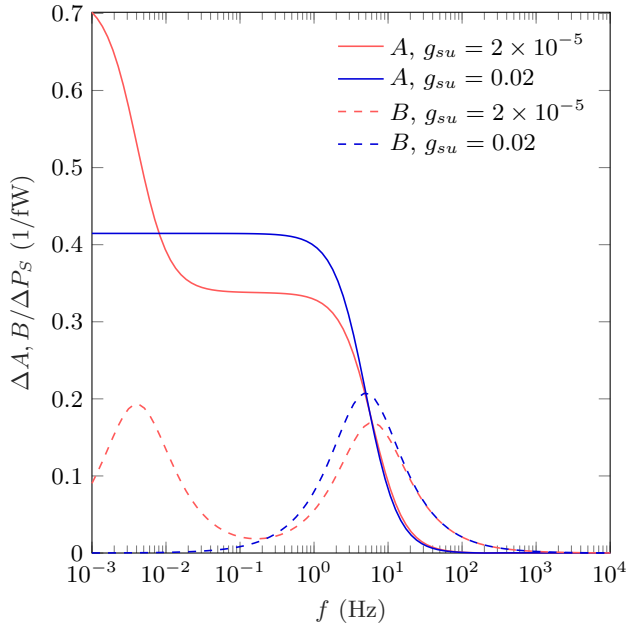}
  \caption{Responsivity of $IQ$ output in principal directions $A$ (solid) and $B$ (dashed) to a small change in signal power $P_S$, at selected values of thermal isolation of the device from the bath, as a function of the modulation frequency $f$ of the signal power change. The change in the response bandwidth is due to the changes in the dynamical response time of the device as the effective thermal conductance changes.
  Al resonator, with zero detuning of readout frequency, and $P_R = \SI{0.2}{fW}$.}
  \label{fig:15}
\end{figure}

Since the responsivity $\propto G\tu{total}^{-1}$,
decreasing the effective thermal conductance of the device by increasing the thermal isolation from the bath also increases the magnitude of the responsivity. \Cref{fig:15} shows this as well as the corresponding decrease in the device response bandwidth. When the substrate-bath conductivity $G_L$ is decreased by a factor of \num{E3}, so it is comparable to the superconductor quasiparticle-phonon conductance $G_I$, the device response bandwidth shows two separate time constants; in this case one is associated with cooling of the quasiparticles into the superconductor and substrate phonons, and the other is the cooling of the entire device into the bath.

\subsection{Noise and NEP}

Noise is introduced into the model in a way similar to equilibrium analysis -- we consider a Langevin source such that the internal energy of each of the thermal elements fluctuates according to its temperature
  \begin{equation}
  \langle \Delta\vc{U}(t)\Delta\vc{U}(t)^\dagger \rangle = k_B \mat{C_0} \vdot \mat{T_0} \vdot \mat{T_0} \;,
  \end{equation}
where $\qty{\mat{T_0}}_{mn} = \qty{\vc{T_0}}_m \delta^{}_{mn}$ is a diagonal matrix constructed from $\vc{T_0}$.
This leads to a noise power vector $\Delta\vc{P}$ added to the power flow vector $\vc{q}$, with spectral power density~\cite{Thomas2014}
  \begin{equation}
    \langle \Delta\vc{P}(f) \Delta\vc{P}(f)^\dagger \rangle = 2 k_B \left( \mat{G_0} \vdot \mat{T_0} \vdot \mat{T_0} + \mat{T_0} \vdot \mat{T_0} \vdot \mat{G_0}^\dagger \right) \; \label{eq:noise}
  \end{equation}
This expression can be compared to the expression for noise power limited by phonon fluctuations in a thermal conductance $G$, well known in bolometer theory~\cite{Goldie2011}, of $\langle \Delta{}P(f) \Delta{}P(f)^* \rangle = 4 k_B T^2 G$.
We note that in the $IQ$ output, this noise corresponds only to noise in one direction in the $IQ$ plane, as the only source is temperature fluctuations in $T_{qp}$.

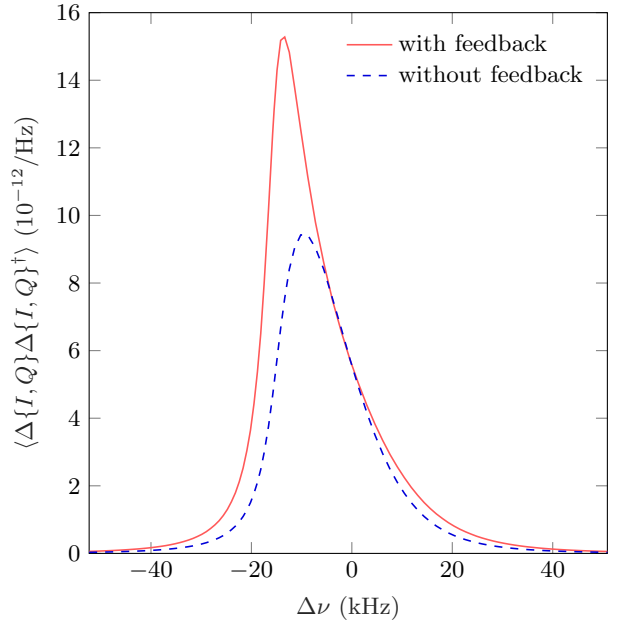
\begin{figure}[htb]
  \centering
  \tikzsetnextfilename{el_figure_18_noise_detuning_feedback}
%
%
\begin{tikzpicture}

\begin{axis}[%
width=0.951\figurewidth,
height=\figureheight,
at={(0\figurewidth,0\figureheight)},
scale only axis,
xmin=-52.3114525680542,
xmax=50.8829405050278,
xlabel style={font=\color{white!15!black}},
xlabel={$\Delta\nu$ (\si{kHz})},
ymin=0,
ymax=16,
ylabel style={font=\color{white!15!black}},
ylabel={$\langle \Delta{}\{I,Q\} \Delta{}\{I,Q\}^\dagger \rangle$ (\num{e-12}/Hz)},
legend style={legend cell align=left, align=left, fill=none, draw=none}
]
\addplot [color=white!35!red]
  table[row sep=crcr]{%
-52.3114525680542	0.0594042535245232\\
-51.3469808359146	0.0638160472900874\\
-50.3807222776413	0.068630905278665\\
-49.4162190666199	0.0743507045387216\\
-48.4535711669922	0.0801139651945589\\
-47.4928868103027	0.0864144936859945\\
-46.5342916097641	0.0938991215280699\\
-45.5779147033691	0.101505543229435\\
-44.6238974866867	0.110544928190016\\
-43.6723876218796	0.120532917386541\\
-42.7235590524673	0.13077694894082\\
-41.777586022377	0.142961738452943\\
-40.8346609649658	0.156498749985726\\
-39.894994099617	0.171564334320975\\
-38.9588148746491	0.189516892607295\\
-38.0263783121109	0.208409815274025\\
-37.0979399347305	0.229576686962709\\
-36.1738061542511	0.254871589093604\\
-35.2595088214874	0.283354628244842\\
-34.3459388046265	0.315747589496294\\
-33.4379286251068	0.352567178404598\\
-32.5359467763901	0.394542480880114\\
-31.6405044517517	0.44253930676274\\
-30.752189332962	0.500523665720407\\
-29.8716252622604	0.564253420011787\\
-28.9942606210709	0.641957467748699\\
-28.1297122573853	0.736498578852258\\
-27.2746973762512	0.842705727731468\\
-26.4299482326508	0.967500745493606\\
-25.6025496921539	1.12035285093658\\
-24.7838314266205	1.30281827482899\\
-23.9731747093201	1.53105875841286\\
-23.1803024358749	1.7954820155683\\
-22.411022187233	2.12430525508753\\
-21.65129641819	2.52709668442592\\
-20.9060832567215	3.03887468044754\\
-20.1895220499039	3.64008906873198\\
-19.4874004745483	4.42858658883393\\
-18.7997560386658	5.37565221155078\\
-18.1398224277496	6.50518679445207\\
-17.4657120494843	7.98560084111825\\
-16.5895724411011	10.1907676656212\\
-15.7650143117905	12.4352022240794\\
-15.4828045911789	13.1173050951808\\
-14.8937410945892	14.3015033128891\\
-14.1720927343369	15.1806855222967\\
-13.3692919721603	15.2799659375076\\
-12.4544202337265	14.8287584751343\\
-11.399892580986	13.7892747334457\\
-10.1734951105118	12.533117795187\\
-8.78393710231781	11.1534560409821\\
-7.24736009311676	9.79704641293946\\
-5.59389046287537	8.59889735944906\\
-3.87399905014038	7.50111858287625\\
-2.12698771286011	6.57476350523461\\
-0.38716445350647	5.74107440124983\\
1.32060879135132	5.02464104267433\\
2.98084754943848	4.38661136417456\\
4.58578451824188	3.83115771845485\\
6.13324117469788	3.34578321819503\\
7.62464794826508	2.9209401964299\\
9.05974321842194	2.5599427805969\\
10.4517645683289	2.2422595088418\\
11.8001243066788	1.96084345155329\\
13.1096071243286	1.71964387470822\\
14.3847393512726	1.50602981404351\\
15.6247146606445	1.32563935080484\\
16.8446795415878	1.16683831752543\\
18.0408227424622	1.03249798985331\\
19.2160905199051	0.910055408918502\\
20.3730422906876	0.806621873282264\\
21.5139253702164	0.715620162951472\\
22.6406734962463	0.638680689213051\\
23.7500669727325	0.570873174722032\\
24.8589732570648	0.507848000326833\\
25.9519212436676	0.455856170803851\\
27.0369163560867	0.409883649697065\\
28.1142680978775	0.367908436881829\\
29.1848167228699	0.332377963565828\\
30.2493008918762	0.300639164848977\\
31.3083624591827	0.272252186827862\\
32.3625698251724	0.246828863238875\\
33.4124269924164	0.224029750397347\\
34.4583815326691	0.203557225520618\\
35.5008159513473	0.186202929252722\\
36.5400888996124	0.17051956055198\\
37.5764953927994	0.155425833362007\\
38.6103191766739	0.142630421475837\\
39.6418048448563	0.13101972144403\\
40.6711712236404	0.120473648953573\\
41.6986152877808	0.111540228641355\\
42.7243046731949	0.102755762520609\\
43.7484074697495	0.0947463007337753\\
44.7710630598068	0.0879668338837916\\
45.7923964252472	0.0812488016074707\\
46.812525229454	0.075565191722011\\
47.8315566196442	0.0703376046519175\\
48.8495794706345	0.0651207442462399\\
49.8666808643341	0.060710105952107\\
50.8829405050278	0.0566410819176088\\
};
\addlegendentry{with feedback}

\addplot [color=black!15!blue, dashed]
  table[row sep=crcr]{%
-52.3114525680542	0.0315760024393508\\
-51.3469808359146	0.0338920231618799\\
-50.3807222776413	0.0364160470519806\\
-49.4162190666199	0.0394080725646002\\
-48.4535711669922	0.0424173648987285\\
-47.4928868103027	0.0457003092679297\\
-46.5342916097641	0.0495912213290277\\
-45.5779147033691	0.0535362398534303\\
-44.6238974866867	0.0582113249164369\\
-43.6723876218796	0.063361683315703\\
-42.7235590524673	0.0686281501039951\\
-41.777586022377	0.0748712944226362\\
-40.8346609649658	0.08178096261266\\
-39.894994099617	0.0894410557056456\\
-38.9588148746491	0.0985262661336136\\
-38.0263783121109	0.10804176917155\\
-37.0979399347305	0.118648192574385\\
-36.1738061542511	0.131249117463491\\
-35.2595088214874	0.145347163784976\\
-34.3459388046265	0.16126768501269\\
-33.4379286251068	0.17922776972034\\
-32.5359467763901	0.19953164579909\\
-31.6405044517517	0.222538692410698\\
-30.752189332962	0.250045381633998\\
-29.8716252622604	0.279948294632518\\
-28.9942606210709	0.315959539941241\\
-28.1297122573853	0.359150970030198\\
-27.2746973762512	0.406942959446237\\
-26.4299482326508	0.462183323374933\\
-25.6025496921539	0.528592567471257\\
-24.7838314266205	0.606263495115385\\
-23.9731747093201	0.701183404777762\\
-23.1803024358749	0.808543910544715\\
-22.411022187233	0.938488335529938\\
-21.65129641819	1.09323356952726\\
-20.9060832567215	1.28376591426746\\
-20.1895220499039	1.50135452894745\\
-19.4874004745483	1.7772557398936\\
-18.7997560386658	2.10175708310683\\
-18.1398224277496	2.48326523854677\\
-17.4657120494843	2.98068740481928\\
-16.5895724411011	3.76430915205815\\
-15.7650143117905	4.68233687783543\\
-15.4828045911789	5.02066869138907\\
-14.8937410945892	5.75561891532989\\
-14.1720927343369	6.66395560177213\\
-13.3692919721603	7.55005941957242\\
-12.4544202337265	8.40567235152291\\
-11.399892580986	9.03623461634545\\
-10.1734951105118	9.42388293392166\\
-8.78393710231781	9.42382900647664\\
-7.24736009311676	9.03788144864579\\
-5.59389046287537	8.40539663865455\\
-3.87399905014038	7.54956216056872\\
-2.12698771286011	6.66341318226222\\
-0.38716445350647	5.75382808911568\\
1.32060879135132	4.92163019105003\\
2.98084754943848	4.16470851645\\
4.58578451824188	3.51082337991234\\
6.13324117469788	2.95411345684103\\
7.62464794826508	2.48464168086297\\
9.05974321842194	2.10165589758053\\
10.4517645683289	1.77944942992477\\
11.8001243066788	1.50718779133881\\
13.1096071243286	1.2833360685358\\
14.3847393512726	1.09362865923426\\
15.6247146606445	0.939363722546866\\
16.8446795415878	0.808445844831223\\
18.0408227424622	0.701163022673093\\
19.2160905199051	0.606551703876987\\
20.3730422906876	0.528742284716082\\
21.5139253702164	0.462040476430122\\
22.6406734962463	0.406860454071702\\
23.7500669727325	0.35924920571645\\
24.8589732570648	0.315917321670547\\
25.9519212436676	0.280720488056551\\
27.0369163560867	0.250115921049508\\
28.1142680978775	0.222588337948261\\
29.1848167228699	0.199569622590114\\
30.2493008918762	0.179257150999045\\
31.3083624591827	0.161291852656912\\
32.3625698251724	0.145367508059527\\
33.4124269924164	0.131220898661859\\
34.4583815326691	0.118628210829913\\
35.5008159513473	0.108028635428502\\
36.5400888996124	0.0985175526552143\\
37.5764953927994	0.0894359935692648\\
38.6103191766739	0.0817786081188355\\
39.6418048448563	0.0748707267688316\\
40.6711712236404	0.0686288261810808\\
41.6986152877808	0.0633629598377435\\
42.7243046731949	0.0582131919645849\\
43.7484074697495	0.0535383019463803\\
44.7710630598068	0.0495934068367174\\
45.7923964252472	0.0457031979948363\\
46.812525229454	0.0424201752585827\\
47.8315566196442	0.039410674912154\\
48.8495794706345	0.0364186970291365\\
49.8666808643341	0.0338947624931906\\
50.8829405050278	0.0315727638904337\\
};
\addlegendentry{without feedback}

\end{axis}
\end{tikzpicture}%
  \caption{Noise squared spectral density in $IQ$ output due to temperature fluctuations, with (solid) and without (dashed) the effects of electrothermal feedback, as a function of readout frequency detuning $\Delta\nu = \nu_r - \nu_0$. In this model only temperature fluctuation noise is included, so noise peaks when quasiparticle temperature $T_{qp}$ is at a maximum. Electrothermal feedback decreases the effective thermal conductance and so decreases the noise in the power flows, but also increases the responsivity such that the noise in the $IQ$ output increases in magnitude.
  With $P_R = \SI{25}{fW}$.}
  \label{fig:18}
\end{figure}

This fluctuation in quasiparticle effective temperature is, microscopically, the generation-recombination noise of the superconductor quasiparticles~\cite{Visser2012}. However, there are three important points: this expression should be evaluated at the effective temperature of the quasiparticles $T_{qp}$, which may be significantly higher than the bath temperature $T_{b}$; the overall noise is scaled by $1/\eta_r$, as in addition to quasiparticle recombination, there is also quasiparticle-phonon scattering contributing to a fluctuation in the effective temperature; and finally, the electrothermal feedback further changes the effective $G$ and so changes the magnitude of the observed noise. The overall effect of the first two factors is to increase the noise estimates over a calculation assuming $T_{qp} = T_b$, while the electrothermal feedback then reduces it slightly. Overall, we find for typical conditions the noise power is higher than an estimate assuming $T_{qp} = T_b$.
We must also distinguish the noise in the power flows \cref{eq:noise}, which reduces due to electrothermal feedback, compared to the consequent noise in the $IQ$ output, which includes the electrical and thermal responsivity~\cite{Thomas2014} and is increased overall by the electrothermal feedback. The noise in the $IQ$ output is plotted in \cref{fig:18}, with and without the effects of electrothermal feedback. Assuming a uniform spectrum for the temperature fluctuations as for Langevin sources, the bandwidth of the noise in the $IQ$ output is determined by the same dynamical time constants as the responsivity.

\begin{figure}[htb]
  \centering
  \tikzsetnextfilename{el_figure_14_nep_feedback}
  \input{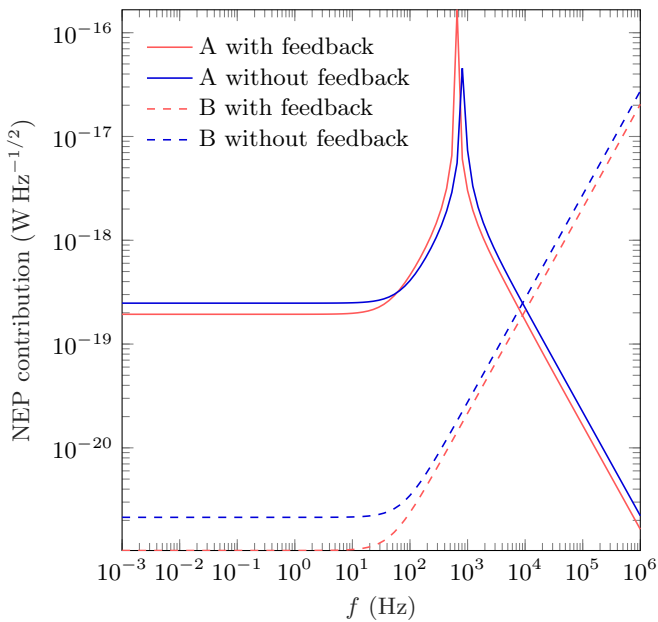}
  \caption{Noise-equivalent signal power (NEP) in $IQ$ output, contribution from intrinsic temperature fluctuations, along the directions of A (solid) and B (dashed) from \cref{eq:AB} in the $IQ$ plane, with and without the effects of electrothermal feedback. NEP peaks at the limits of the response bandwidth of the device. Electrothermal feedback reduces the effective thermal conductance and so reduces the NEP.
  With zero detuning of readout frequency, and $P_R = \SI{2}{fW}$.}
  \label{fig:14}
\end{figure}

Noise Equivalent Power (NEP) is a common figure of merit for detectors; it is defined in this case for signal power into a detector as 
\begin{equation}
\text{NEP}_x = \frac{\sqrt{\langle \Delta{}x \Delta{}x^\dagger \rangle}}{\dv*{x}{P_S}}  \;,
\end{equation}
where $x$ is the output as experimentally measured, for example the $IQ$ output. Lower NEPs correspond to more sensitive detectors; KIDs were originally proposed to operate at NEPs below $\SI{E-19}{W.Hz^{-1/2}}$~\cite{Day2003}.
The NEP contribution from the intrinsic noise of the device is plotted in \cref{fig:14}, as a function of the modulation frequency of the signal power. Since in this model the axis of thermal noise in the $IQ$ plane and the axes of maximum and minimum $IQ$ responsivity to signal power do not need to align, the NEP contribution is split up into two components along $A$ and $B$. The NEP contribution is flat within the response bandwidth of the device, then increases outside that bandwidth.
In practice, there would also be other noise sources, for example from the readout system or photon noise, which would also limit the achievable NEP. At low frequencies in particular, there can be a significant $1/f$ component to the noise~\cite{Baselmans2008}, constraining the performance of very narrow bandwidth devices with high thermal isolation or $Q$-factors.
Overall we find the contribution of intrinsic thermal noise to NEP, as measured in the $IQ$ output, increases with the effective total thermal conductance as $\text{NEP} \propto G\tu{total}^{1/2}$. This means that the electrothermal feedback reducing $G\tu{total}$ has a net benefit, reducing the NEP, as shown in \cref{fig:14}.
For comparison, the NEP of a bolometer limited by phonon noise in a thermal conductance~\cite{Osman2014} $G$ is $\sqrt{4 k_B T^2 G}$ in equlibrium.

\section{Conclusions} \label{sec:conclusions}

Both the large-signal operating point and small-signal device behaviour of a Kinetic Inductance Detector may be affected by readout power heating of the superconductor quasiparticles -- electrothermal feedback. Implementing a general, extensible model of a typical Kinetic Inductance Detector allows us to investigate the effect on device behaviour.
The steady-state operating point, for a given readout frequency and applied readout power, clearly shows the effects of readout power heating, which can be increased by increasing the thermal isolation of the device. This readout power heating and electrothermal feedback lead to nonlinear resonances, asymmetric about the resonant frequency; at high readout powers, we observe hysteresis and multiple stable operating points.
Two regimes of operation are possible; one in which the superconductor quasiparticle-phonon thermal conductance is the limiting factor, and so the superconductor phonons and substrate phonons are approximately at the bath temperature; and one in which the substrate-bath thermal conductance is the limiting factor, and in steady-state operation the superconductor phonons and substrate phonons may be significantly above the bath temperature. We find that when quasiparticle heating is significant, the nonlinear functional form of the quasiparticle-phonon thermal conductance appears to isolate the quasiparticle temperature from other parameters, such as substrate heating power and bath temperature. This effect will be explored and quantified in further work.

Linearising the behaviour of the device about the steady-state operating point, we consider the ellipses created in the $IQ$ plane by sinusoidal modulations of the temperature-independent parameters, and define the optimal responsivity of the device by finding the two principal axes. Following this readout strategy allows us to show that the maximum responsivity is at the resonant frequency when readout power heating is not significant, but slightly below when heating is significant.
The device output responsivity to signal (pair-breaking) power is inversely proportional to the effective overall conductance from the quasiparticles to the bath $G\tu{total}^{-1}$, as is the noise squared in the $IQ$ output due to temperature fluctuations. This leads to a Noise Equivalent Power $\propto G\tu{total}^{1/2}$.
Therefore increasing the thermal isolation of the device while keeping quasiparticle temperature constant (for example by reducing readout power), decreasing the effective thermal conductance from the superconductor quasiparticles to the bath decreases the NEP of the device. For typical operating conditions we also find the contribution of electrothermal feedback to be positive, reducing the overall thermal conductance, and so helping to reduce the NEP of the device.
However, there are tradeoffs between reducing thermal conductance and the dynamic range and bandwidth, as well as between the increased noise from a higher quasiparticle temperature, and responsivity to bath temperature fluctuations. For highly thermally isolated devices, we also expect nonlinearities will become prominent at lower absorbed powers. These conclusions may inform device design of the next generation of KIDs attempting to reach NEPs of \SI{E-20}{W.Hz^{-1/2}}, required by a range of far-infrared astronomy applications~\cite{Goldie2011}.

The extensible nature of this model leaves many possible avenues for further work, including investigating how background signal loading affects device performance; including more noise sources, for example from the readout system or phase noise, and calculating their effect on NEP and the optimal readout scheme; and including other common nonlinear phenomena affecting KIDs, such as Two-Level Systems (TLS) which as significant at low readout powers, and the intrinsic nonlinearity of kinetic inductance due to the change in Cooper pair states in superconductors carrying high currents. Although not considered here, the model is also applicable to devices intended to detect pulsed signal power, and so may be useful in calculating expected pulse shapes for high-energy photon detection applications.

\bibliography{paper_electrothermal}

\begin{thebibliography}{36}%
\makeatletter
\providecommand \@ifxundefined [1]{%
 \@ifx{#1\undefined}
}%
\providecommand \@ifnum [1]{%
 \ifnum #1\expandafter \@firstoftwo
 \else \expandafter \@secondoftwo
 \fi
}%
\providecommand \@ifx [1]{%
 \ifx #1\expandafter \@firstoftwo
 \else \expandafter \@secondoftwo
 \fi
}%
\providecommand \natexlab [1]{#1}%
\providecommand \enquote  [1]{``#1''}%
\providecommand \bibnamefont  [1]{#1}%
\providecommand \bibfnamefont [1]{#1}%
\providecommand \citenamefont [1]{#1}%
\providecommand \href@noop [0]{\@secondoftwo}%
\providecommand \href [0]{\begingroup \@sanitize@url \@href}%
\providecommand \@href[1]{\@@startlink{#1}\@@href}%
\providecommand \@@href[1]{\endgroup#1\@@endlink}%
\providecommand \@sanitize@url [0]{\catcode `\\12\catcode `\$12\catcode
  `\&12\catcode `\#12\catcode `\^12\catcode `\_12\catcode `\%12\relax}%
\providecommand \@@startlink[1]{}%
\providecommand \@@endlink[0]{}%
\providecommand \url  [0]{\begingroup\@sanitize@url \@url }%
\providecommand \@url [1]{\endgroup\@href {#1}{\urlprefix }}%
\providecommand \urlprefix  [0]{URL }%
\providecommand \Eprint [0]{\href }%
\providecommand \doibase [0]{http://dx.doi.org/}%
\providecommand \selectlanguage [0]{\@gobble}%
\providecommand \bibinfo  [0]{\@secondoftwo}%
\providecommand \bibfield  [0]{\@secondoftwo}%
\providecommand \translation [1]{[#1]}%
\providecommand \BibitemOpen [0]{}%
\providecommand \bibitemStop [0]{}%
\providecommand \bibitemNoStop [0]{.\EOS\space}%
\providecommand \EOS [0]{\spacefactor3000\relax}%
\providecommand \BibitemShut  [1]{\csname bibitem#1\endcsname}%
\let\auto@bib@innerbib\@empty
\bibitem [{\citenamefont {Day}\ \emph {et~al.}(2003)\citenamefont {Day},
  \citenamefont {LeDuc}, \citenamefont {Mazin}, \citenamefont {Vayonakis},\
  and\ \citenamefont {Zmuidzinas}}]{Day2003}%
  \BibitemOpen
  \bibfield  {author} {\bibinfo {author} {\bibfnamefont {P.~K.}\ \bibnamefont
  {Day}}, \bibinfo {author} {\bibfnamefont {H.~G.}\ \bibnamefont {LeDuc}},
  \bibinfo {author} {\bibfnamefont {B.~A.}\ \bibnamefont {Mazin}}, \bibinfo
  {author} {\bibfnamefont {A.}~\bibnamefont {Vayonakis}}, \ and\ \bibinfo
  {author} {\bibfnamefont {J.}~\bibnamefont {Zmuidzinas}},\ }\href {\doibase
  10.1038/nature02037} {\bibfield  {journal} {\bibinfo  {journal} {Nature}\
  }\textbf {\bibinfo {volume} {425}},\ \bibinfo {pages} {817} (\bibinfo {year}
  {2003})}\BibitemShut {NoStop}%
\bibitem [{\citenamefont {Vardulakis}\ \emph {et~al.}(2008)\citenamefont
  {Vardulakis}, \citenamefont {Withington}, \citenamefont {Goldie},\ and\
  \citenamefont {Glowacka}}]{Vardulakis2008}%
  \BibitemOpen
  \bibfield  {author} {\bibinfo {author} {\bibfnamefont {G.}~\bibnamefont
  {Vardulakis}}, \bibinfo {author} {\bibfnamefont {S.}~\bibnamefont
  {Withington}}, \bibinfo {author} {\bibfnamefont {D.~J.}\ \bibnamefont
  {Goldie}}, \ and\ \bibinfo {author} {\bibfnamefont {D.~M.}\ \bibnamefont
  {Glowacka}},\ }\href {\doibase 10.1088/0957-0233/19/1/015509} {\bibfield
  {journal} {\bibinfo  {journal} {Meas. Sci. Technol.}\ }\textbf {\bibinfo
  {volume} {19}},\ \bibinfo {pages} {015509} (\bibinfo {year}
  {2008})}\BibitemShut {NoStop}%
\bibitem [{\citenamefont {Cruciani}\ \emph {et~al.}(2016)\citenamefont
  {Cruciani}, \citenamefont {Bellini}, \citenamefont {Cardani}, \citenamefont
  {Casali}, \citenamefont {Castellano}, \citenamefont {Colantoni},
  \citenamefont {Coppolecchia}, \citenamefont {Cosmelli}, \citenamefont
  {D'Addabbo}, \citenamefont {{Di Domizio}}, \citenamefont {Martinez},
  \citenamefont {Tomei},\ and\ \citenamefont {Vignati}}]{Cruciani2016}%
  \BibitemOpen
  \bibfield  {author} {\bibinfo {author} {\bibfnamefont {A.}~\bibnamefont
  {Cruciani}}, \bibinfo {author} {\bibfnamefont {F.}~\bibnamefont {Bellini}},
  \bibinfo {author} {\bibfnamefont {L.}~\bibnamefont {Cardani}}, \bibinfo
  {author} {\bibfnamefont {N.}~\bibnamefont {Casali}}, \bibinfo {author}
  {\bibfnamefont {M.~G.}\ \bibnamefont {Castellano}}, \bibinfo {author}
  {\bibfnamefont {I.}~\bibnamefont {Colantoni}}, \bibinfo {author}
  {\bibfnamefont {A.}~\bibnamefont {Coppolecchia}}, \bibinfo {author}
  {\bibfnamefont {C.}~\bibnamefont {Cosmelli}}, \bibinfo {author}
  {\bibfnamefont {A.}~\bibnamefont {D'Addabbo}}, \bibinfo {author}
  {\bibfnamefont {S.}~\bibnamefont {{Di Domizio}}}, \bibinfo {author}
  {\bibfnamefont {M.}~\bibnamefont {Martinez}}, \bibinfo {author}
  {\bibfnamefont {C.}~\bibnamefont {Tomei}}, \ and\ \bibinfo {author}
  {\bibfnamefont {M.}~\bibnamefont {Vignati}},\ }\href {\doibase
  10.1007/s10909-016-1574-0} {\bibfield  {journal} {\bibinfo  {journal} {J. Low
  Temp. Phys.}\ }\textbf {\bibinfo {volume} {184}},\ \bibinfo {pages} {859}
  (\bibinfo {year} {2016})}\BibitemShut {NoStop}%
\bibitem [{\citenamefont {DiCarlo}\ \emph {et~al.}(2010)\citenamefont
  {DiCarlo}, \citenamefont {Reed}, \citenamefont {Sun}, \citenamefont
  {Johnson}, \citenamefont {Chow}, \citenamefont {Gambetta}, \citenamefont
  {Frunzio}, \citenamefont {Girvin}, \citenamefont {Devoret},\ and\
  \citenamefont {Schoelkopf}}]{DiCarlo2010}%
  \BibitemOpen
  \bibfield  {author} {\bibinfo {author} {\bibfnamefont {L.}~\bibnamefont
  {DiCarlo}}, \bibinfo {author} {\bibfnamefont {M.~D.}\ \bibnamefont {Reed}},
  \bibinfo {author} {\bibfnamefont {L.}~\bibnamefont {Sun}}, \bibinfo {author}
  {\bibfnamefont {B.~R.}\ \bibnamefont {Johnson}}, \bibinfo {author}
  {\bibfnamefont {J.~M.}\ \bibnamefont {Chow}}, \bibinfo {author}
  {\bibfnamefont {J.~M.}\ \bibnamefont {Gambetta}}, \bibinfo {author}
  {\bibfnamefont {L.}~\bibnamefont {Frunzio}}, \bibinfo {author} {\bibfnamefont
  {S.~M.}\ \bibnamefont {Girvin}}, \bibinfo {author} {\bibfnamefont {M.~H.}\
  \bibnamefont {Devoret}}, \ and\ \bibinfo {author} {\bibfnamefont {R.~J.}\
  \bibnamefont {Schoelkopf}},\ }\href {\doibase 10.1038/nature09416} {\bibfield
   {journal} {\bibinfo  {journal} {Nature}\ }\textbf {\bibinfo {volume}
  {467}},\ \bibinfo {pages} {574} (\bibinfo {year} {2010})}\BibitemShut
  {NoStop}%
\bibitem [{\citenamefont {Hofheinz}\ \emph {et~al.}(2008)\citenamefont
  {Hofheinz}, \citenamefont {Weig}, \citenamefont {Ansmann}, \citenamefont
  {Bialczak}, \citenamefont {Lucero}, \citenamefont {Neeley}, \citenamefont
  {O'Connell}, \citenamefont {Wang}, \citenamefont {Martinis},\ and\
  \citenamefont {Cleland}}]{Hofheinz2008}%
  \BibitemOpen
  \bibfield  {author} {\bibinfo {author} {\bibfnamefont {M.}~\bibnamefont
  {Hofheinz}}, \bibinfo {author} {\bibfnamefont {E.~M.}\ \bibnamefont {Weig}},
  \bibinfo {author} {\bibfnamefont {M.}~\bibnamefont {Ansmann}}, \bibinfo
  {author} {\bibfnamefont {R.~C.}\ \bibnamefont {Bialczak}}, \bibinfo {author}
  {\bibfnamefont {E.}~\bibnamefont {Lucero}}, \bibinfo {author} {\bibfnamefont
  {M.}~\bibnamefont {Neeley}}, \bibinfo {author} {\bibfnamefont {A.~D.}\
  \bibnamefont {O'Connell}}, \bibinfo {author} {\bibfnamefont {H.}~\bibnamefont
  {Wang}}, \bibinfo {author} {\bibfnamefont {J.~M.}\ \bibnamefont {Martinis}},
  \ and\ \bibinfo {author} {\bibfnamefont {A.~N.}\ \bibnamefont {Cleland}},\
  }\href {\doibase 10.1038/nature07136} {\bibfield  {journal} {\bibinfo
  {journal} {Nature}\ }\textbf {\bibinfo {volume} {454}},\ \bibinfo {pages}
  {310} (\bibinfo {year} {2008})}\BibitemShut {NoStop}%
\bibitem [{\citenamefont {Schoelkopf}\ and\ \citenamefont
  {Girvin}(2008)}]{Schoelkopf2008}%
  \BibitemOpen
  \bibfield  {author} {\bibinfo {author} {\bibfnamefont {R.~J.}\ \bibnamefont
  {Schoelkopf}}\ and\ \bibinfo {author} {\bibfnamefont {S.~M.}\ \bibnamefont
  {Girvin}},\ }\href {\doibase 10.1038/451664a} {\bibfield  {journal} {\bibinfo
   {journal} {Nature}\ }\textbf {\bibinfo {volume} {451}},\ \bibinfo {pages}
  {664} (\bibinfo {year} {2008})}\BibitemShut {NoStop}%
\bibitem [{\citenamefont {Irwin}\ and\ \citenamefont
  {Lehnert}(2004)}]{Irwin2004}%
  \BibitemOpen
  \bibfield  {author} {\bibinfo {author} {\bibfnamefont {K.~D.}\ \bibnamefont
  {Irwin}}\ and\ \bibinfo {author} {\bibfnamefont {K.~W.}\ \bibnamefont
  {Lehnert}},\ }\href {\doibase 10.1063/1.1791733} {\bibfield  {journal}
  {\bibinfo  {journal} {Appl. Phys. Lett.}\ }\textbf {\bibinfo {volume} {85}},\
  \bibinfo {pages} {2107} (\bibinfo {year} {2004})}\BibitemShut {NoStop}%
\bibitem [{\citenamefont {Zmuidzinas}(2012)}]{Zmuidzinas2012}%
  \BibitemOpen
  \bibfield  {author} {\bibinfo {author} {\bibfnamefont {J.}~\bibnamefont
  {Zmuidzinas}},\ }\href {\doibase 10.1146/annurev-conmatphys-020911-125022}
  {\bibfield  {journal} {\bibinfo  {journal} {Annu. Rev. Condens. Matter
  Phys.}\ }\textbf {\bibinfo {volume} {3}},\ \bibinfo {pages} {169} (\bibinfo
  {year} {2012})}\BibitemShut {NoStop}%
\bibitem [{\citenamefont {de~Visser}\ \emph {et~al.}(2010)\citenamefont
  {de~Visser}, \citenamefont {Withington},\ and\ \citenamefont
  {Goldie}}]{DeVisser2010}%
  \BibitemOpen
  \bibfield  {author} {\bibinfo {author} {\bibfnamefont {P.~J.}\ \bibnamefont
  {de~Visser}}, \bibinfo {author} {\bibfnamefont {S.}~\bibnamefont
  {Withington}}, \ and\ \bibinfo {author} {\bibfnamefont {D.~J.}\ \bibnamefont
  {Goldie}},\ }\href {\doibase 10.1063/1.3517152} {\bibfield  {journal}
  {\bibinfo  {journal} {J. Appl. Phys.}\ }\textbf {\bibinfo {volume} {108}},\
  \bibinfo {pages} {114504} (\bibinfo {year} {2010})}\BibitemShut {NoStop}%
\bibitem [{\citenamefont {Thompson}\ \emph {et~al.}(2013)\citenamefont
  {Thompson}, \citenamefont {Withington}, \citenamefont {Goldie},\ and\
  \citenamefont {Thomas}}]{Thompson2013}%
  \BibitemOpen
  \bibfield  {author} {\bibinfo {author} {\bibfnamefont {S.~E.}\ \bibnamefont
  {Thompson}}, \bibinfo {author} {\bibfnamefont {S.}~\bibnamefont
  {Withington}}, \bibinfo {author} {\bibfnamefont {D.~J.}\ \bibnamefont
  {Goldie}}, \ and\ \bibinfo {author} {\bibfnamefont {C.~N.}\ \bibnamefont
  {Thomas}},\ }\href {\doibase 10.1088/0953-2048/26/9/095009} {\bibfield
  {journal} {\bibinfo  {journal} {Supercond. Sci. Technol.}\ }\textbf {\bibinfo
  {volume} {26}},\ \bibinfo {pages} {095009} (\bibinfo {year}
  {2013})}\BibitemShut {NoStop}%
\bibitem [{\citenamefont {Goldie}\ and\ \citenamefont
  {Withington}(2013)}]{Goldie2013}%
  \BibitemOpen
  \bibfield  {author} {\bibinfo {author} {\bibfnamefont {D.~J.}\ \bibnamefont
  {Goldie}}\ and\ \bibinfo {author} {\bibfnamefont {S.}~\bibnamefont
  {Withington}},\ }\href {\doibase 10.1088/0953-2048/26/1/015004} {\bibfield
  {journal} {\bibinfo  {journal} {Supercond. Sci. Technol.}\ }\textbf {\bibinfo
  {volume} {26}},\ \bibinfo {pages} {015004} (\bibinfo {year} {2013})},\
  \Eprint {http://arxiv.org/abs/1208.0685} {arXiv:1208.0685} \BibitemShut
  {NoStop}%
\bibitem [{\citenamefont {Guruswamy}\ \emph {et~al.}(2015)\citenamefont
  {Guruswamy}, \citenamefont {Goldie},\ and\ \citenamefont
  {Withington}}]{Guruswamy2015b}%
  \BibitemOpen
  \bibfield  {author} {\bibinfo {author} {\bibfnamefont {T.}~\bibnamefont
  {Guruswamy}}, \bibinfo {author} {\bibfnamefont {D.~J.}\ \bibnamefont
  {Goldie}}, \ and\ \bibinfo {author} {\bibfnamefont {S.}~\bibnamefont
  {Withington}},\ }\href {\doibase 10.1088/0953-2048/28/5/054002} {\bibfield
  {journal} {\bibinfo  {journal} {Supercond. Sci. Technol.}\ }\textbf {\bibinfo
  {volume} {28}},\ \bibinfo {pages} {054002} (\bibinfo {year} {2015})},\
  \Eprint {http://arxiv.org/abs/1501.01831} {arXiv:1501.01831} \BibitemShut
  {NoStop}%
\bibitem [{\citenamefont {Guruswamy}\ \emph {et~al.}(2016)\citenamefont
  {Guruswamy}, \citenamefont {Goldie},\ and\ \citenamefont
  {Withington}}]{Guruswamy2016}%
  \BibitemOpen
  \bibfield  {author} {\bibinfo {author} {\bibfnamefont {T.}~\bibnamefont
  {Guruswamy}}, \bibinfo {author} {\bibfnamefont {D.~J.}\ \bibnamefont
  {Goldie}}, \ and\ \bibinfo {author} {\bibfnamefont {S.}~\bibnamefont
  {Withington}},\ }\href {\doibase 10.1088/0953-2048/29/4/045011} {\bibfield
  {journal} {\bibinfo  {journal} {Supercond. Sci. Technol.}\ }\textbf {\bibinfo
  {volume} {29}},\ \bibinfo {pages} {045011} (\bibinfo {year} {2016})},\
  \Eprint {http://arxiv.org/abs/1602.06860} {arXiv:1602.06860} \BibitemShut
  {NoStop}%
\bibitem [{\citenamefont {Thomas}\ \emph {et~al.}(2015)\citenamefont {Thomas},
  \citenamefont {Withington},\ and\ \citenamefont {Goldie}}]{Thomas2014}%
  \BibitemOpen
  \bibfield  {author} {\bibinfo {author} {\bibfnamefont {C.~N.}\ \bibnamefont
  {Thomas}}, \bibinfo {author} {\bibfnamefont {S.}~\bibnamefont {Withington}},
  \ and\ \bibinfo {author} {\bibfnamefont {D.~J.}\ \bibnamefont {Goldie}},\
  }\href {\doibase 10.1088/0953-2048/28/4/045012} {\bibfield  {journal}
  {\bibinfo  {journal} {Supercond. Sci. Technol.}\ }\textbf {\bibinfo {volume}
  {28}},\ \bibinfo {pages} {045012} (\bibinfo {year} {2015})},\ \Eprint
  {http://arxiv.org/abs/1411.1565} {arXiv:1411.1565} \BibitemShut {NoStop}%
\bibitem [{\citenamefont {Goldie}\ \emph {et~al.}(2011)\citenamefont {Goldie},
  \citenamefont {Velichko}, \citenamefont {Glowacka},\ and\ \citenamefont
  {Withington}}]{Goldie2011}%
  \BibitemOpen
  \bibfield  {author} {\bibinfo {author} {\bibfnamefont {D.~J.}\ \bibnamefont
  {Goldie}}, \bibinfo {author} {\bibfnamefont {A.~V.}\ \bibnamefont
  {Velichko}}, \bibinfo {author} {\bibfnamefont {D.~M.}\ \bibnamefont
  {Glowacka}}, \ and\ \bibinfo {author} {\bibfnamefont {S.}~\bibnamefont
  {Withington}},\ }\href {\doibase 10.1063/1.3561432} {\bibfield  {journal}
  {\bibinfo  {journal} {J. Appl. Phys.}\ }\textbf {\bibinfo {volume} {109}},\
  \bibinfo {pages} {084507} (\bibinfo {year} {2011})}\BibitemShut {NoStop}%
\bibitem [{\citenamefont {Osman}\ \emph {et~al.}(2014)\citenamefont {Osman},
  \citenamefont {Withington}, \citenamefont {Goldie},\ and\ \citenamefont
  {Glowacka}}]{Osman2014}%
  \BibitemOpen
  \bibfield  {author} {\bibinfo {author} {\bibfnamefont {D.}~\bibnamefont
  {Osman}}, \bibinfo {author} {\bibfnamefont {S.}~\bibnamefont {Withington}},
  \bibinfo {author} {\bibfnamefont {D.~J.}\ \bibnamefont {Goldie}}, \ and\
  \bibinfo {author} {\bibfnamefont {D.~M.}\ \bibnamefont {Glowacka}},\ }\href
  {\doibase 10.1063/1.4893019} {\bibfield  {journal} {\bibinfo  {journal} {J.
  Appl. Phys.}\ }\textbf {\bibinfo {volume} {116}},\ \bibinfo {pages} {064506}
  (\bibinfo {year} {2014})}\BibitemShut {NoStop}%
\bibitem [{\citenamefont {Quaranta}\ \emph {et~al.}(2013)\citenamefont
  {Quaranta}, \citenamefont {Cecil}, \citenamefont {Gades}, \citenamefont
  {Mazin},\ and\ \citenamefont {Miceli}}]{Quaranta2013}%
  \BibitemOpen
  \bibfield  {author} {\bibinfo {author} {\bibfnamefont {O.}~\bibnamefont
  {Quaranta}}, \bibinfo {author} {\bibfnamefont {T.~W.}\ \bibnamefont {Cecil}},
  \bibinfo {author} {\bibfnamefont {L.}~\bibnamefont {Gades}}, \bibinfo
  {author} {\bibfnamefont {B.}~\bibnamefont {Mazin}}, \ and\ \bibinfo {author}
  {\bibfnamefont {A.}~\bibnamefont {Miceli}},\ }\href {\doibase
  10.1088/0953-2048/26/10/105021} {\bibfield  {journal} {\bibinfo  {journal}
  {Supercond. Sci. Technol.}\ }\textbf {\bibinfo {volume} {26}},\ \bibinfo
  {pages} {105021} (\bibinfo {year} {2013})}\BibitemShut {NoStop}%
\bibitem [{\citenamefont {Timofeev}\ \emph {et~al.}(2014)\citenamefont
  {Timofeev}, \citenamefont {Vesterinen}, \citenamefont {Helist{\"{o}}},
  \citenamefont {Gr{\"{o}}nberg}, \citenamefont {Hassel},\ and\ \citenamefont
  {Luukanen}}]{Timofeev2014}%
  \BibitemOpen
  \bibfield  {author} {\bibinfo {author} {\bibfnamefont {A.~V.}\ \bibnamefont
  {Timofeev}}, \bibinfo {author} {\bibfnamefont {V.}~\bibnamefont
  {Vesterinen}}, \bibinfo {author} {\bibfnamefont {P.}~\bibnamefont
  {Helist{\"{o}}}}, \bibinfo {author} {\bibfnamefont {L.}~\bibnamefont
  {Gr{\"{o}}nberg}}, \bibinfo {author} {\bibfnamefont {J.}~\bibnamefont
  {Hassel}}, \ and\ \bibinfo {author} {\bibfnamefont {A.}~\bibnamefont
  {Luukanen}},\ }\href {\doibase 10.1088/0953-2048/27/2/025002} {\bibfield
  {journal} {\bibinfo  {journal} {Supercond. Sci. Technol.}\ }\textbf {\bibinfo
  {volume} {27}},\ \bibinfo {pages} {025002} (\bibinfo {year}
  {2014})}\BibitemShut {NoStop}%
\bibitem [{\citenamefont {Lindeman}\ \emph {et~al.}(2014)\citenamefont
  {Lindeman}, \citenamefont {Bonetti}, \citenamefont {Bumble}, \citenamefont
  {Day}, \citenamefont {Eom}, \citenamefont {Holmes},\ and\ \citenamefont
  {Kleinsasser}}]{Lindeman2014b}%
  \BibitemOpen
  \bibfield  {author} {\bibinfo {author} {\bibfnamefont {M.~A.}\ \bibnamefont
  {Lindeman}}, \bibinfo {author} {\bibfnamefont {J.~A.}\ \bibnamefont
  {Bonetti}}, \bibinfo {author} {\bibfnamefont {B.}~\bibnamefont {Bumble}},
  \bibinfo {author} {\bibfnamefont {P.~K.}\ \bibnamefont {Day}}, \bibinfo
  {author} {\bibfnamefont {B.~H.}\ \bibnamefont {Eom}}, \bibinfo {author}
  {\bibfnamefont {W.~A.}\ \bibnamefont {Holmes}}, \ and\ \bibinfo {author}
  {\bibfnamefont {A.~W.}\ \bibnamefont {Kleinsasser}},\ }\href {\doibase
  10.1063/1.4884437} {\bibfield  {journal} {\bibinfo  {journal} {J. Appl.
  Phys.}\ }\textbf {\bibinfo {volume} {115}},\ \bibinfo {pages} {234509}
  (\bibinfo {year} {2014})}\BibitemShut {NoStop}%
\bibitem [{\citenamefont {Lindeman}\ \emph {et~al.}(2013)\citenamefont
  {Lindeman}, \citenamefont {Khosropanah},\ and\ \citenamefont
  {Hijmering}}]{Lindeman2013}%
  \BibitemOpen
  \bibfield  {author} {\bibinfo {author} {\bibfnamefont {M.~A.}\ \bibnamefont
  {Lindeman}}, \bibinfo {author} {\bibfnamefont {P.}~\bibnamefont
  {Khosropanah}}, \ and\ \bibinfo {author} {\bibfnamefont {R.~A.}\ \bibnamefont
  {Hijmering}},\ }\href {\doibase 10.1063/1.4790146} {\bibfield  {journal}
  {\bibinfo  {journal} {J. Appl. Phys.}\ }\textbf {\bibinfo {volume} {113}},\
  \bibinfo {pages} {074502} (\bibinfo {year} {2013})}\BibitemShut {NoStop}%
\bibitem [{\citenamefont {Lindeman}(2014)}]{Lindeman2014}%
  \BibitemOpen
  \bibfield  {author} {\bibinfo {author} {\bibfnamefont {M.~A.}\ \bibnamefont
  {Lindeman}},\ }\href {\doibase 10.1063/1.4890018} {\bibfield  {journal}
  {\bibinfo  {journal} {J. Appl. Phys.}\ }\textbf {\bibinfo {volume} {116}},\
  \bibinfo {pages} {024506} (\bibinfo {year} {2014})}\BibitemShut {NoStop}%
\bibitem [{\citenamefont {Cardani}\ \emph {et~al.}(2015)\citenamefont
  {Cardani}, \citenamefont {Colantoni}, \citenamefont {Cruciani}, \citenamefont
  {{Di Domizio}}, \citenamefont {Vignati}, \citenamefont {Bellini},
  \citenamefont {Casali}, \citenamefont {Castellano}, \citenamefont
  {Coppolecchia}, \citenamefont {Cosmelli},\ and\ \citenamefont
  {Tomei}}]{Cardani2015}%
  \BibitemOpen
  \bibfield  {author} {\bibinfo {author} {\bibfnamefont {L.}~\bibnamefont
  {Cardani}}, \bibinfo {author} {\bibfnamefont {I.}~\bibnamefont {Colantoni}},
  \bibinfo {author} {\bibfnamefont {A.}~\bibnamefont {Cruciani}}, \bibinfo
  {author} {\bibfnamefont {S.}~\bibnamefont {{Di Domizio}}}, \bibinfo {author}
  {\bibfnamefont {M.}~\bibnamefont {Vignati}}, \bibinfo {author} {\bibfnamefont
  {F.}~\bibnamefont {Bellini}}, \bibinfo {author} {\bibfnamefont
  {N.}~\bibnamefont {Casali}}, \bibinfo {author} {\bibfnamefont {M.~G.}\
  \bibnamefont {Castellano}}, \bibinfo {author} {\bibfnamefont
  {A.}~\bibnamefont {Coppolecchia}}, \bibinfo {author} {\bibfnamefont
  {C.}~\bibnamefont {Cosmelli}}, \ and\ \bibinfo {author} {\bibfnamefont
  {C.}~\bibnamefont {Tomei}},\ }\href {\doibase 10.1063/1.4929977} {\bibfield
  {journal} {\bibinfo  {journal} {Appl. Phys. Lett.}\ }\textbf {\bibinfo
  {volume} {107}},\ \bibinfo {pages} {093508} (\bibinfo {year}
  {2015})}\BibitemShut {NoStop}%
\bibitem [{\citenamefont {Gao}(2008)}]{Gao2008a}%
  \BibitemOpen
  \bibfield  {author} {\bibinfo {author} {\bibfnamefont {J.}~\bibnamefont
  {Gao}},\ }\emph {\bibinfo {title} {{The Physics of Superconducting Microwave
  Resonators}}},\ \href
  {http://resolver.caltech.edu/CaltechETD:etd-06092008-235549} {Ph.D. thesis},\
  \bibinfo  {school} {California Institute of Technology} (\bibinfo {year}
  {2008})\BibitemShut {NoStop}%
\bibitem [{\citenamefont {Mattis}\ and\ \citenamefont
  {Bardeen}(1958)}]{Mattis1958}%
  \BibitemOpen
  \bibfield  {author} {\bibinfo {author} {\bibfnamefont {D.}~\bibnamefont
  {Mattis}}\ and\ \bibinfo {author} {\bibfnamefont {J.}~\bibnamefont
  {Bardeen}},\ }\href {\doibase 10.1103/PhysRev.111.412} {\bibfield  {journal}
  {\bibinfo  {journal} {Phys. Rev.}\ }\textbf {\bibinfo {volume} {111}},\
  \bibinfo {pages} {412} (\bibinfo {year} {1958})}\BibitemShut {NoStop}%
\bibitem [{\citenamefont {Tinkham}(2004)}]{Tinkham2004}%
  \BibitemOpen
  \bibfield  {author} {\bibinfo {author} {\bibfnamefont {M.}~\bibnamefont
  {Tinkham}},\ }\href@noop {} {\emph {\bibinfo {title} {{Introduction to
  Superconductivity}}}}\ (\bibinfo  {publisher} {Dover Publications},\ \bibinfo
  {year} {2004})\ p.\ \bibinfo {pages} {454}\BibitemShut {NoStop}%
\bibitem [{\citenamefont {Phillips}(1987)}]{Phillips1987}%
  \BibitemOpen
  \bibfield  {author} {\bibinfo {author} {\bibfnamefont {W.~A.}\ \bibnamefont
  {Phillips}},\ }\href {\doibase 10.1088/0034-4885/50/12/003} {\bibfield
  {journal} {\bibinfo  {journal} {Reports Prog. Phys.}\ }\textbf {\bibinfo
  {volume} {50}},\ \bibinfo {pages} {1657} (\bibinfo {year}
  {1987})}\BibitemShut {NoStop}%
\bibitem [{\citenamefont {Rostem}\ \emph
  {et~al.}(2008{\natexlab{a}})\citenamefont {Rostem}, \citenamefont {Glowacka},
  \citenamefont {Goldie},\ and\ \citenamefont {Withington}}]{Rostem2008a}%
  \BibitemOpen
  \bibfield  {author} {\bibinfo {author} {\bibfnamefont {K.}~\bibnamefont
  {Rostem}}, \bibinfo {author} {\bibfnamefont {D.~M.}\ \bibnamefont
  {Glowacka}}, \bibinfo {author} {\bibfnamefont {D.~J.}\ \bibnamefont
  {Goldie}}, \ and\ \bibinfo {author} {\bibfnamefont {S.}~\bibnamefont
  {Withington}},\ }in\ \href {\doibase 10.1117/12.787372} {\emph {\bibinfo
  {booktitle} {Proc. SPIE}}},\ Vol.\ \bibinfo {volume} {7020}\ (\bibinfo {year}
  {2008})\ p.\ \bibinfo {pages} {70200L}\BibitemShut {NoStop}%
\bibitem [{\citenamefont {Chang}\ and\ \citenamefont
  {Scalapino}(1977)}]{Chang1977}%
  \BibitemOpen
  \bibfield  {author} {\bibinfo {author} {\bibfnamefont {J.-J.}\ \bibnamefont
  {Chang}}\ and\ \bibinfo {author} {\bibfnamefont {D.~J.}\ \bibnamefont
  {Scalapino}},\ }\href {\doibase 10.1103/PhysRevB.15.2651} {\bibfield
  {journal} {\bibinfo  {journal} {Phys. Rev. B}\ }\textbf {\bibinfo {volume}
  {15}},\ \bibinfo {pages} {2651} (\bibinfo {year} {1977})}\BibitemShut
  {NoStop}%
\bibitem [{\citenamefont {Chang}\ and\ \citenamefont
  {Scalapino}(1978)}]{Chang1978}%
  \BibitemOpen
  \bibfield  {author} {\bibinfo {author} {\bibfnamefont {J.-J.}\ \bibnamefont
  {Chang}}\ and\ \bibinfo {author} {\bibfnamefont {D.~J.}\ \bibnamefont
  {Scalapino}},\ }\href {\doibase 10.1007/BF00116228} {\bibfield  {journal}
  {\bibinfo  {journal} {J. Low Temp. Phys.}\ }\textbf {\bibinfo {volume}
  {31}},\ \bibinfo {pages} {1} (\bibinfo {year} {1978})}\BibitemShut {NoStop}%
\bibitem [{\citenamefont {Maisi}\ \emph {et~al.}(2013)\citenamefont {Maisi},
  \citenamefont {Lotkhov}, \citenamefont {Kemppinen}, \citenamefont {Heimes},
  \citenamefont {Muhonen},\ and\ \citenamefont {Pekola}}]{Maisi2013}%
  \BibitemOpen
  \bibfield  {author} {\bibinfo {author} {\bibfnamefont {V.~F.}\ \bibnamefont
  {Maisi}}, \bibinfo {author} {\bibfnamefont {S.~V.}\ \bibnamefont {Lotkhov}},
  \bibinfo {author} {\bibfnamefont {A.}~\bibnamefont {Kemppinen}}, \bibinfo
  {author} {\bibfnamefont {A.}~\bibnamefont {Heimes}}, \bibinfo {author}
  {\bibfnamefont {J.~T.}\ \bibnamefont {Muhonen}}, \ and\ \bibinfo {author}
  {\bibfnamefont {J.~P.}\ \bibnamefont {Pekola}},\ }\href {\doibase
  10.1103/PhysRevLett.111.147001} {\bibfield  {journal} {\bibinfo  {journal}
  {Phys. Rev. Lett.}\ }\textbf {\bibinfo {volume} {111}},\ \bibinfo {pages}
  {147001} (\bibinfo {year} {2013})}\BibitemShut {NoStop}%
\bibitem [{\citenamefont {Rostem}\ \emph
  {et~al.}(2008{\natexlab{b}})\citenamefont {Rostem}, \citenamefont {Glowacka},
  \citenamefont {Goldie},\ and\ \citenamefont {Withington}}]{Rostem2008}%
  \BibitemOpen
  \bibfield  {author} {\bibinfo {author} {\bibfnamefont {K.}~\bibnamefont
  {Rostem}}, \bibinfo {author} {\bibfnamefont {D.}~\bibnamefont {Glowacka}},
  \bibinfo {author} {\bibfnamefont {D.~J.}\ \bibnamefont {Goldie}}, \ and\
  \bibinfo {author} {\bibfnamefont {S.}~\bibnamefont {Withington}},\ }\href
  {\doibase 10.1007/s10909-007-9701-6} {\bibfield  {journal} {\bibinfo
  {journal} {J. Low Temp. Phys.}\ }\textbf {\bibinfo {volume} {151}},\ \bibinfo
  {pages} {76} (\bibinfo {year} {2008}{\natexlab{b}})}\BibitemShut {NoStop}%
\bibitem [{\citenamefont {Kelley}(1995)}]{Kelley1995}%
  \BibitemOpen
  \bibfield  {author} {\bibinfo {author} {\bibfnamefont {C.~T.}\ \bibnamefont
  {Kelley}},\ }\href {\doibase 10.1137/1.9781611970944} {\emph {\bibinfo
  {title} {{Iterative Methods for Linear and Nonlinear Equations}}}},\ \bibinfo
  {series} {Frontiers in Applied Mathematics}, Vol.~\bibinfo {volume} {16}\
  (\bibinfo  {publisher} {SIAM},\ \bibinfo {address} {Philadelphia, PA},\
  \bibinfo {year} {1995})\BibitemShut {NoStop}%
\bibitem [{\citenamefont {Swenson}\ \emph {et~al.}(2013)\citenamefont
  {Swenson}, \citenamefont {Day}, \citenamefont {Eom}, \citenamefont {Leduc},
  \citenamefont {Llombart}, \citenamefont {McKenney}, \citenamefont
  {Noroozian},\ and\ \citenamefont {Zmuidzinas}}]{Swenson2013}%
  \BibitemOpen
  \bibfield  {author} {\bibinfo {author} {\bibfnamefont {L.~J.}\ \bibnamefont
  {Swenson}}, \bibinfo {author} {\bibfnamefont {P.~K.}\ \bibnamefont {Day}},
  \bibinfo {author} {\bibfnamefont {B.~H.}\ \bibnamefont {Eom}}, \bibinfo
  {author} {\bibfnamefont {H.~G.}\ \bibnamefont {Leduc}}, \bibinfo {author}
  {\bibfnamefont {N.}~\bibnamefont {Llombart}}, \bibinfo {author}
  {\bibfnamefont {C.~M.}\ \bibnamefont {McKenney}}, \bibinfo {author}
  {\bibfnamefont {O.}~\bibnamefont {Noroozian}}, \ and\ \bibinfo {author}
  {\bibfnamefont {J.}~\bibnamefont {Zmuidzinas}},\ }\href {\doibase
  10.1063/1.4794808} {\bibfield  {journal} {\bibinfo  {journal} {J. Appl.
  Phys.}\ }\textbf {\bibinfo {volume} {113}},\ \bibinfo {pages} {104501}
  (\bibinfo {year} {2013})}\BibitemShut {NoStop}%
\bibitem [{\citenamefont {Semenov}\ \emph {et~al.}(2016)\citenamefont
  {Semenov}, \citenamefont {Devyatov}, \citenamefont {de~Visser},\ and\
  \citenamefont {Klapwijk}}]{Semenov2016}%
  \BibitemOpen
  \bibfield  {author} {\bibinfo {author} {\bibfnamefont {A.~V.}\ \bibnamefont
  {Semenov}}, \bibinfo {author} {\bibfnamefont {I.~A.}\ \bibnamefont
  {Devyatov}}, \bibinfo {author} {\bibfnamefont {P.~J.}\ \bibnamefont
  {de~Visser}}, \ and\ \bibinfo {author} {\bibfnamefont {T.~M.}\ \bibnamefont
  {Klapwijk}},\ }\href {\doibase 10.1103/PhysRevLett.117.047002} {\bibfield
  {journal} {\bibinfo  {journal} {Phys. Rev. Lett.}\ }\textbf {\bibinfo
  {volume} {117}},\ \bibinfo {pages} {047002} (\bibinfo {year}
  {2016})}\BibitemShut {NoStop}%
\bibitem [{\citenamefont {de~Visser}\ \emph {et~al.}(2012)\citenamefont
  {de~Visser}, \citenamefont {Baselmans}, \citenamefont {Diener}, \citenamefont
  {Yates}, \citenamefont {Endo},\ and\ \citenamefont {Klapwijk}}]{Visser2012}%
  \BibitemOpen
  \bibfield  {author} {\bibinfo {author} {\bibfnamefont {P.~J.}\ \bibnamefont
  {de~Visser}}, \bibinfo {author} {\bibfnamefont {J.~J.~A.}\ \bibnamefont
  {Baselmans}}, \bibinfo {author} {\bibfnamefont {P.}~\bibnamefont {Diener}},
  \bibinfo {author} {\bibfnamefont {S.~J.~C.}\ \bibnamefont {Yates}}, \bibinfo
  {author} {\bibfnamefont {A.}~\bibnamefont {Endo}}, \ and\ \bibinfo {author}
  {\bibfnamefont {T.~M.}\ \bibnamefont {Klapwijk}},\ }\href {\doibase
  10.1007/s10909-012-0519-5} {\bibfield  {journal} {\bibinfo  {journal} {J. Low
  Temp. Phys.}\ }\textbf {\bibinfo {volume} {167}},\ \bibinfo {pages} {335}
  (\bibinfo {year} {2012})}\BibitemShut {NoStop}%
\bibitem [{\citenamefont {Baselmans}\ \emph {et~al.}(2008)\citenamefont
  {Baselmans}, \citenamefont {Yates}, \citenamefont {Barends}, \citenamefont
  {Lankwarden}, \citenamefont {Gao}, \citenamefont {Hoevers},\ and\
  \citenamefont {Klapwijk}}]{Baselmans2008}%
  \BibitemOpen
  \bibfield  {author} {\bibinfo {author} {\bibfnamefont {J.}~\bibnamefont
  {Baselmans}}, \bibinfo {author} {\bibfnamefont {S.~J.~C.}\ \bibnamefont
  {Yates}}, \bibinfo {author} {\bibfnamefont {R.}~\bibnamefont {Barends}},
  \bibinfo {author} {\bibfnamefont {Y.~J.~Y.}\ \bibnamefont {Lankwarden}},
  \bibinfo {author} {\bibfnamefont {J.~R.}\ \bibnamefont {Gao}}, \bibinfo
  {author} {\bibfnamefont {H.}~\bibnamefont {Hoevers}}, \ and\ \bibinfo
  {author} {\bibfnamefont {T.~M.}\ \bibnamefont {Klapwijk}},\ }\href {\doibase
  10.1007/s10909-007-9684-3} {\bibfield  {journal} {\bibinfo  {journal} {J. Low
  Temp. Phys.}\ }\textbf {\bibinfo {volume} {151}},\ \bibinfo {pages} {524}
  (\bibinfo {year} {2008})}\BibitemShut {NoStop}%
\end{thebibliography}%
\end{document}